\documentclass[a4paper,useAMS,usenatbib]{mnras}
\pdfoutput=1 

\usepackage{graphicx}
\usepackage{setspace}
\usepackage{natbib}
\usepackage{color}
\usepackage{amsmath,amssymb}
\usepackage{times}

\usepackage{hyperref}[]

\title[Nitrogen enrichment and clustered star formation at the dawn of the Galaxy]{Nitrogen enrichment and clustered star formation at the dawn of the Galaxy}

\author[Belokurov \& Kravtsov]{Vasily  Belokurov$^{1}$\thanks{E-mail:vasily@ast.cam.ac.uk} and Andrey Kravtsov$^{2,3,4}$\thanks{E-mail:kravtsov@uchicago.edu}\\
  $^1$Institute of Astronomy, Madingley Rd, Cambridge, CB3 0HA, UK\\ 
  $^2$Department of Astronomy and Astrophysics, The University of Chicago, Chicago, IL 60637 USA\\
  $^3$Kavli Institute for Cosmological Physics, The University of Chicago, Chicago, IL 60637 USA\\
  $^4$Enrico Fermi Institute, The University of Chicago, Chicago, IL 60637}

\begin{document}
\defcitealias{Aurora}{BK22}
%\date{August 2015}
%\pagerange{\pageref{firstpage}--\pageref{lastpage}} \pubyear{2015}

\maketitle

\label{firstpage}

\begin{abstract}
Anomalously high nitrogen-to-oxygen abundance ratios [N/O] are observed in globular clusters (GCs), among the field stars of the Milky Way (MW), and
even in the gas in a $z\approx 11$ galaxy. 
Using data from the APOGEE Data Release 17 and the {\it Gaia} Data Release 3, we 
present several independent lines of evidence that most of the MW's high-[N/O] stars were born in situ in massive bound clusters during the early, pre-disk evolution of the Galaxy. Specifically, we show that distributions of metallicity [Fe/H], energy, the angular momentum $L_z$, and distance of the low-metallicity high-[N/O] stars match the corresponding distributions of stars of the {\it Aurora} population {\it and} of the in-situ GCs. We also show that the fraction of in-situ field high-[N/O] stars, $f_{\rm N/O}$, increases rapidly with decreasing metallicity. During epochs when metallicity evolves from $\rm [Fe/H]=-1.5$ to $\rm [Fe/H]=-0.9$, the Galaxy spins up and transitions from a turbulent Aurora state to a coherently rotating disk. This transformation is accompanied by many qualitative changes. In particular, we show that high N/O abundances similar to those observed in GN-z11 were common before the spin-up ($\rm [Fe/H]\lesssim -1.5$) when up to $\approx 50\%-70\%$ of the in-situ stars formed in massive bound clusters. The dramatic drop of $f_{\rm N/O}$ at  $\rm [Fe/H]\gtrsim -0.9$ indicates that after the disk emerges the fraction of stars forming in massive bound clusters decreases by two orders of magnitude. 

\end{abstract}


\section{Introduction}
\label{sec:intro}

The dawn of the Universal galaxy assembly can now be explored in tantalizing detail via direct high-resolution infrared JWST observations of high redshift galaxies \citep[see, e.g.,][]{Donnan2023, Harikane2023, Finkelstein2023,Robertson2023}. Thanks to the combined power of JWST's NIRCam and NIRSpec instruments, galaxies beyond $z=8$ are revealed to be small, dense, metal-poor, and actively star-forming \citep[see, e.g.,][]{Ono2022,Tacchella2023,Robertson2023,Bowens2023,Curtis-Lake2023}. 

In the Milky Way (MW), independent constraints on the physics of high-$z$ star formation can be obtained using Galactic archaeology which interprets properties of individual ancient, low-metallicity stars from large surveys. The two approaches are complementary: direct observations of high-$z$ galaxies probe the star formation and state of the interstellar medium enriched by the nucleosynthetic products from the first generations of massive stars, while Galactic archeology explores properties of the surviving low-mass stars that formed in the low-metallicity environment of the MW's progenitor.

With the NIRSpec instrument, the brightest of the high-$z$ galaxies are now amenable to unprecedented levels of scrutiny, going far beyond a simple spectroscopic redshift measurement. One such example is GN-z11, identified previously with the HST and {\it Spitzer} \citep[][]{Oesch2016}. The follow-up observations with NIRSpec reported measurements of oxygen, carbon, and neon lines, as well as unusually prominent levels of nitrogen emission, indicating high relative [N/O] abundance at moderately low oxygen abundance [O/H] \citep[][]{Bunker2023}. Using these measurements  \citet{Cameron2023} estimated the relative nitrogen abundance in GN-z11 to be $\log({\rm N}/{\rm O})\gtrsim -0.25$, which is much higher than this ratio in the Sun, i.e. $\log({\rm N}/{\rm O})_{\odot}= -0.86$ \citep[see][]{Lodders2019}. 

In the Galaxy's stellar populations such nitrogen over-abundance is rare but exists: stars with high [N/O] ratio are numerous in Globular Clusters \citep[see, e.g.,][]{Bastian_Lardo2018, Gratton2019, Miloni_Marino2022}. This similarity in the abundance patterns and possible connection between enrichment pathways in local globular clusters and GN-z11 has been recently pointed out in several studies \citep[][]{Cameron2023, Senchyna2023, Charbonnel2023}.

Nitrogen over-abundance in the MW GCs is so blatant that it has been used routinely as a chemical fingerprint to identify field stars that were born in clusters \citep[see, e.g.,][]{Martell_Grebel2010,Carollo2013,Martell2016,Schiavon2017,Fernandez_Trincado2017,Tang2019,Horta2021,Phillips2022}. The cluster origin of such field stars is evidenced by their correlations and anti-correlations of abundances of different chemical elements similar to those observed in GCs: e.g., depleted [O/Fe] and [Mg/Fe] and enhanced [Al/Fe] \citep[see e.g.][]{Lind2015,Schiavon2017,Fernandez_Trincado2020b,Horta2021}. Curiously, not all high-[N/O] stars show the rest of the GC-specific chemical pattern. For example, a population of N-enhanced giants discovered by \citet{Fernandez_Trincado2020} in the Magellanic Clouds is consistent with the typical MW field in other projections of the chemical abundance space. 

Given that the GC-born stars are straightforward to pick out in the field, a number of studies estimated the overall fraction of Galactic stellar mass contributed by clusters under various assumptions. For example, most studies agree that the overall observed fraction of stars with high [N/O] (hereafter high-[N/O] stars) at $\rm [Fe/H]<-1$ is rather low,  $\approx 2\%-3\%$, but somewhat higher estimates can be obtained depending on the threshold in nitrogen enrichment, the metallicity range and the location in the Galaxy used \citep[see, e.g.,][]{Martell2016, Schiavon2017, Koch2019, Horta2021}. To convert the observed high-[N/O] fraction into the total stellar mass born in clusters assumptions are made as to the initial mass of the star clusters disrupted by $z=0$. 

Another factor is the role of the so-called "first population" (1P) GC stars in clusters. 
The 1P stars themselves have chemical abundances indistinguishable from the rest of the Galaxy's field but they are assumed to directly contribute to the anomalous chemistry of the ``second generation'' (2P), which is manifested in N, Na and Al enrichment and C, Mg and O depletion \citep{Bastian_Lardo2018, Gratton2019, Miloni_Marino2022}. Using the hypothesis that the 1P stars could have made up to $\sim$90\% of the initial cluster's mass \citep[e.g.,][]{Dercole2008,Conroy2012,Bastian2013}, the measured $\sim$2\% implies that $\sim$20\%  of the field stars is contributed by clusters \citep[e.g.,][]{Martell2011}. Assuming a more conservative value of the fraction of 1P stars, $f_{\rm 1P}\approx 0.5$ comparable to a typical fraction observed in surviving MW clusters \citep[][]{Milone2017}, the inferred GC contribution to the Galactic halo's stellar mass is reduced considerably to $\sim 5\%$ \citep[see][]{Koch2019}.

Interpretation of such estimates, however, is not straightforward.  In the metallicity range of the high-[N/O] stars, stellar population is a mix of stars brought in by other galaxies (accreted) and stars born in the MW's main progenitor (in-situ). Thus, the fraction of high-[N/O] halo stars is an average of stars born in clusters in all of the progenitor galaxies that contributed to the MW's stellar halo.  To address this, several studies attempted to assign high-[N/O] stars to distinct halo components based, for example, on the [Al/Fe] ratio \citep[][]{Kisku2021,Fernandez_Trincado2022}. However, given the generally anomalous chemical abundances of many of the GC-born stars, including [Al/Fe], such an assignment is bound to be biased. Orbital information has also been used but with inconclusive results \citep[][]{Trincado2020c,Tang2020,Fernandez_Trincado2022}. 

Several studies reported a prominent population of N-rich stars residing in the Milky Way's bulge \citep[see][]{Schiavon2017,Jurassic}. The increased incidence of N-rich stars towards the Galactic centre is also reported in \citet{Horta2021}. Unfortunately, the star's presence in the bulge does not elucidate where and when it formed. Multiple origins remain viable: bulge stars can be part of either accreted or in-situ halo population or even belong to the bar. 

Recently, data from the {\it Gaia} satellite has helped to clarify and systematize the make-up of the Galactic stellar halo. In confirmation of the hypothesis put forward by \citet{Deason2013}, the bulk of the accreted debris within 40 kpc from the Galactic centre appears to be donated via a single, massive and ancient merger event known as the {\it Gaia} Sausage/Enceladus \citep[GS/E,][]{Belokurov2018,Helmi2018}. A sizeable population of the Milky Way GCs was shown to belong to the GS/E progenitor dwarf galaxy by \citet{Myeong2018}. Details of the Galactic GC classification have been considered and re-evaluated many a time since \citep[e.g.,][]{Massari2019, Krujssen2019, Myeong2019,Forbes2020,Callingham2022}. A consensus amongst these works is that many of the GCs formed in situ can be identified based on either their location in the age-metallicity space or by their high angular momentum. Several recent efforts have started to verify this nomenclature via high-resolution spectroscopic studies \citep[e.g.][]{Koch-Hansen2021,McKenzie2022,Monty2023}.

Not surprisingly, across the above classification efforts, the in-situ GCs have lower average energy compared to the accreted ones. Sometimes a fraction of the low-energy GCs is assigned to a separate accretion event \citep[][]{Massari2019, Forbes2020, Krijssen2020, Callingham2022}. It is unclear however what makes these low-energy GCs distinct from the rest of the in-situ clusters. In the original analysis of \citet{Massari2019}, which inspired many of the follow-up works mentioned above, the in-situ GCs were split into two groups, the tightly bound ``bulge'' and the clearly rotating ``disk'' clusters. 

Recently, however, the metal-poor ($\rm [Fe/H]<-1$) portion of the in-situ stellar halo has been shown to have a much wider range of azimuthal velocities compared to the accreted component, at least in the vicinity of the Sun \citep[][hereafter BK22]{Aurora}. 
Discovered through chemical tagging, this component, dubbed {\it Aurora}, is the oldest Milky Way stellar population formed before the Galaxy had a coherently rotating disk.  

Aurora stellar population spans a range of energies, from the lowest levels typical for the stars near the Galactic centre to that of the Sun, but its density beyond the Solar radius falls sharply. The distribution of azimuthal velocities of the Aurora stars is significantly broader than that of the GS/E. However, unlike GS/E's debris which has little net rotation, Aurora has a modest net spin of $\sim50$ km s$^{-1}$. BK22 show that at higher metallicities ($\rm [Fe/H]>-1$) the kinematic behaviour of the ancient MW stars exhibits a clear trend: the azimuthal velocity increases sharply as the Galaxy spins up to become a disk \citep[see also][]{Conroy2022,Rix2022}. 

Aurora's stars also exhibit a large scatter in most elements but in particular in Al, N, O, and Si, i.e. the same elements that are considered as the best GC markers due to their anomalous behaviour. Consequently, BK22 conclude that Aurora's chemistry likely bears signs of a large contribution from massive star clusters, a hypothesis strengthened in the analysis of \citet{Myeong2022}.

Thus there are multiple indications that instead of an agglomeration of numerous fragments of distinct origin, the metal-poor stellar halo inside the Solar radius is dominated by one prominent population formed in-situ at early epochs. Throughout this Paper we refer to this pre-disk component as {\it Aurora} to emphasize its in-situ origin thus accepting that it may contain most or all of the alleged Kraken/Koala/Heracles structure \citep[][]{Krujssen2019,Horta_heracles,Forbes2020}. 

The main reasons to consider such a monolithic, single-origin classification scheme for the bulk of the metal-poor portion in the inner MW halo are twofold. First, detailed high-resolution chemical studies show little difference between Aurora and Kraken/Koala/Heracles \citep[][]{Aurora,Naidu2022,Myeong2022,Horta2023}. Second, a look at the chemo-kinematics of these stars shows a clear continuity in their orbital properties across a wide range of angular momenta, from retrograde to mildly rotating and a wide range of energies, from the most bound, "bulge"-like to approximately Solar \citep[][]{PIGS_I,PIGS_II,Aurora,Conroy2022,Myeong2022,Rix2022}. Consequently, in our study, field stars and Galactic GCs  classified as in-situ have a broader range of total energies and angular momenta than considered previously.

In this paper we aim to consistently identify the in-situ and accreted components of the MW's stellar population in the metallicity range probed by the APOGEE survey and of the MW's population of globular clusters. To this end, we use both the APOGEE measurements of chemical abundances of elements and {\it Gaia} EDR3 measurements of proper motions of stars and globular clusters. We use the resulting classification to estimate the fraction of stars with enhanced nitrogen abundance in the in-situ and accreted populations and the fraction of low-metallicity stars born in bound clusters. Given that we select such stars using the [N/O] ratio we will refer to these stars in the context of this study as high-[N/O] stars. The paper is structured as follows. Section~\ref{sec:data} presents the details of our selection of field stars, high-[N/O] stars and likely GC members. In Section~\ref{sec:res}, we analyze distributions of the selected stellar populations, decipher the origin of the field high-[N/O] stars and estimate the contribution of GC-like objects to star formation in the early Galaxy. We discuss the implications of our inference in Section~\ref{sec:discussion} where we list the salient changes accompanying the MW's transition from the chaotic Aurora state to the stable disk. Section~\ref{sec:conc} lists our conclusions.

\begin{figure*}
  \centering
  \includegraphics[width=0.99\textwidth]{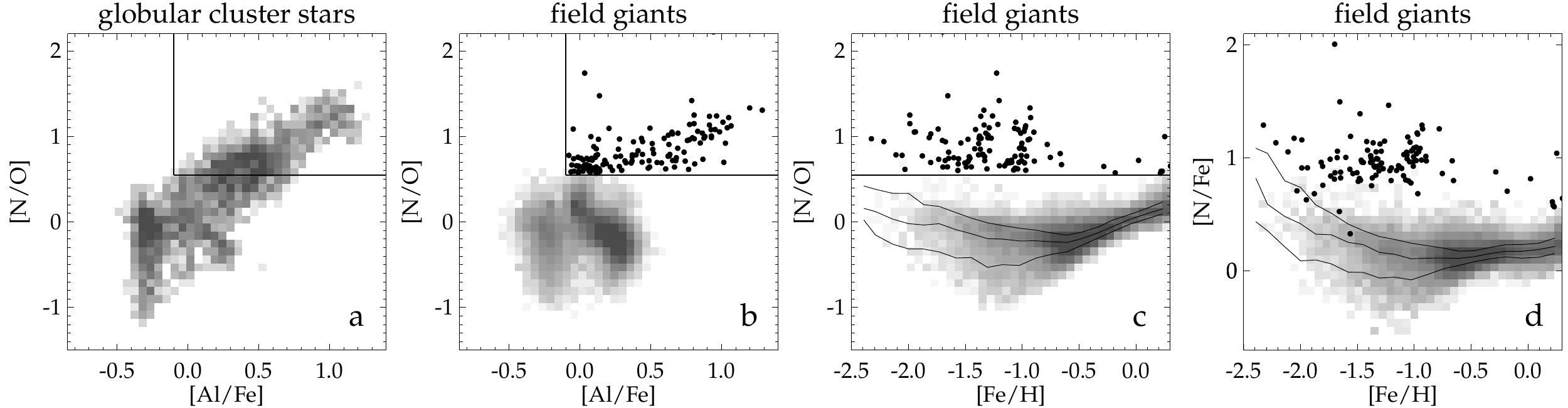}
  \caption[]{Stars with anomalous abundances in APOGEE DR17. {\bf Panel a:} Greyscale shows the density of globular cluster stars in the space of [N/O] and [Al/Fe]. Horizontal (vertical) line is the chosen [N/O] ([Al/Fe]) threshold for the selection of high-[N/O] stars with GC like chemical abundances. {\bf Panel b:} Same as the previous panel but for field giants with $V_\phi<160$ km/s. Small black points are the high-[N/O] stars with GC-like abundances.  {\bf Panel c:} Same as panel b, but for [N/O] vs [Fe/H]. Here and in the next panel, solid black lines show 20th, 50th and 80th percentiles of the abundance ratio distribution as a function of metallicity. {\bf Panel d:} Same as panel c, but for [N/Fe] vs [Fe/H]. Note a stronger upward trend at low metallicity. }
   \label{fig:select}
\end{figure*}
%

%-------------------------------------
\section{Data and sample selection}
\label{sec:data}
%---------------------------------

We use element abundances from the APOGEE Data Release 17 \citep[][]{apogeedr17}, as recorded in the \texttt{allStarLite} catalogue provided on the survey's website. Following BK22, we remove stars with flags:
\verb|STAR_BAD, TEFF_BAD, LOGG_BAD|, \verb|VERY_BRIGHT_NEIGHBOR|, \verb|LOW_SNR, PERSIST_HIGH|, \verb|PERSIST_JUMP_POS|,  \verb|SUSPECT_RV_COMBINATION|, \verb|PERSIST_JUMP_NEG|,  as well as duplicate with \verb|EXTRATARG| flag.
Distances are taken from the AstroNN value-added catalogue \citep[see][]{Leung2019,Mackereth2018}. 
We rely on {\it Gaia} EDR3 proper motions \citep[][]{gaia_edr3,Lindegren2021} and convert observed heliocentric stellar coordinates into the Galactocentric left-handed reference frame, assuming that the Sun is at $X=R_{\odot}=8$ kpc from the Galactic Centre \citep[c.f. a slightly larger value from][]{GRAVITY2022}, and has Galactic $Z_{\odot}=0$. Following \citet{Drimmel2022}, we assume that the Sun's velocity is $v_{\odot}=\{-9.3, 251.5, 8.59\}$ km s$^{-1}$. Total energies $E$ are calculated in a three-component (bulge, disk and DM halo) Galaxy potential identical to that used in \citet{wrinkles}. In what follows, energy is reported in units of $10^5\,\mathrm{km}^2\,\mathrm{s}^{-2}$ and the vertical component of the angular momentum $L_z$ in units of $10^3$ kpc km~s$^{-1}$. 

\subsection{Sample of field red giants with low $V_{\phi}$}
\label{sec:star_sample}

Our base sample of field stars is selected as follows. First, we remove stars within 1.3 degree of all known Galactic satellites (in particular, globular clusters) as well as all objects with  \texttt{PROGRAMNAME=magclouds}. Additionally, we consider only stars within 10 kpc from the Sun that are consistent with being red giants by using the following cuts: $D<10$ kpc, $\log(g)<3$ and $T_{\rm eff}<5300$ K. We also cull stars with tangential velocity errors larger than 50 km s$^{-1}$ and [Fe/H], [N/Fe] and [O/Fe] errors larger than 0.25 dex. Note that removing measurements with large uncertainties can bias [N/O] ratios high at low metallicities. We have checked for the presence of such bias by re-running the entirety of our analysis without a cut on abundance errors and report that any changes in the measurements reported are within their associated uncertainties. Finally, to get rid of the fast-rotating, young stars in the Galaxy's thin disk, we apply a cut on the tangential component of stellar velocity $V_{\phi}<160$ km~s$^{-1}$.  The combination of the above cuts leaves a total of $\sim30,000$ stars. Given the tangential velocity cut applied, stars in this sample are predominantly halo at [Fe/H]$<-1$ and high-$\alpha$ (thick) disk at [Fe/H]$>-1$.

\subsection{Sample of globular cluster stars}

We search the APOGEE DR17 catalog for likely Galactic globular cluster members using the following strategy. For each cluster, likely members are selected within 1.5 times the GC's tidal radius from its centre as reported in the 2010 version of the GC catalog of \citet{Harris2010}. 
%Only stars within 15 km s$^{-1}$ of the reported line-of-sight velocity are kept. Additionally, we require that stellar proper motions are within 1 mas year$^{-1}$ of the {\it Gaia} EDR3-based cluster values reported in \citet{Vasiliev2021}. 
Only stars with cluster membership probabilities above 0.5,  as calculated by \citet{Vasiliev2021}, are kept. Finally, we apply the same $\log(g)$ and $T_{\rm eff}$ cuts as above and retain only those stars whose [Fe/H] in APOGEE is within 0.3 dex of the GC catalog value. This selection procedure yields $\sim4,200$ candidate GC members. When available, we use globular cluster distances  from \citet{Baumgardt2021} and model estimates of their initial masses from \citet{Baumgardt2003}, fractions of the 1st cluster population from \citet{Milone2017}, and GC isochronal ages are from \citet{VdB2013}. Total energy  and $L_z$ angular momentum for each cluster are computed using the same assumptions about the Galaxy as described in the beginning of Section~\ref{sec:data}.

\section{Results} 
\label{sec:res}

\subsection{High-[N/O] field giants in the Milky Way}
\label{sec:no_select}

Our selection of stars with high nitrogen abundances is inspired by previous APOGEE-based studies \citep[e.g.][]{Schiavon2017, Horta2021} with two minor tweaks, as illustrated in Figure~\ref{fig:select}. Instead of applying a cut on [N/Fe], we require {\rm [N/O]}$-\sigma_{\rm [N/O]}>0.55$. Motivation for using [N/O] instead of [N/Fe] is twofold. First, there is a small but noticeable upward [N/Fe] trend with decreasing metallicity (see panel d of Figure~\ref{fig:select}) meaning that a selection based on a single [N/Fe] threshold is not viable. Second, we are looking to compare chemical properties of the MW stars to the extragalactic gas-phase abundances referenced to oxygen. 

We choose the [N/O] threshold i) to match approximately the highest [N/O] ratios in the metal-rich disk population (see panel c of Figure~\ref{fig:select}) and ii) to reach the N/O levels observed in the high-redshift galaxy GN-z11 (see Section~\ref{sec:gnz11}). Note, however, that the exact value of the adopted [N/O] threshold does not affect our results significantly because the derived fractions of GC-born stars are computed self-consistently using a calibration on surviving clusters (see Sections~\ref{sec:trends} and ~\ref{sec:frac_feh}). 

Panel a of Figure~\ref{fig:select} shows the distribution of [N/O] as a function of [Al/Fe] in our sample of GC stars. Stars in GCs exhibit both anomalously high [N/O] and [Al/Fe] ratios. Therefore, for GC-like high-[N/O] stars, we use both the N/O threshold as well as a cut on aluminium, [Al/Fe]$>-0.1$ -- these are shown in panel a with solid black lines. 

Panel b of Figure~\ref{fig:select}  shows the distribution of field giants (selected using the criteria described above) in the space of [N/O] and [Al/Fe]. Although a proportionally much smaller number of high-[N/O] stars is observed in the field, most of them follow a correlation between [N/O] and [Al/Fe] very similar to GC stars. In fact, the overall distribution of GC stars in panel a and the field giants in panel b is quite similar for all values of [N/O] and [Al/Fe]. This is consistent with the conclusion we reached from the analyses presented in this paper that a large fraction of the low-metallicity field stars were born in massive bound clusters.

\begin{figure*}
  \centering
  \includegraphics[width=0.99\textwidth]{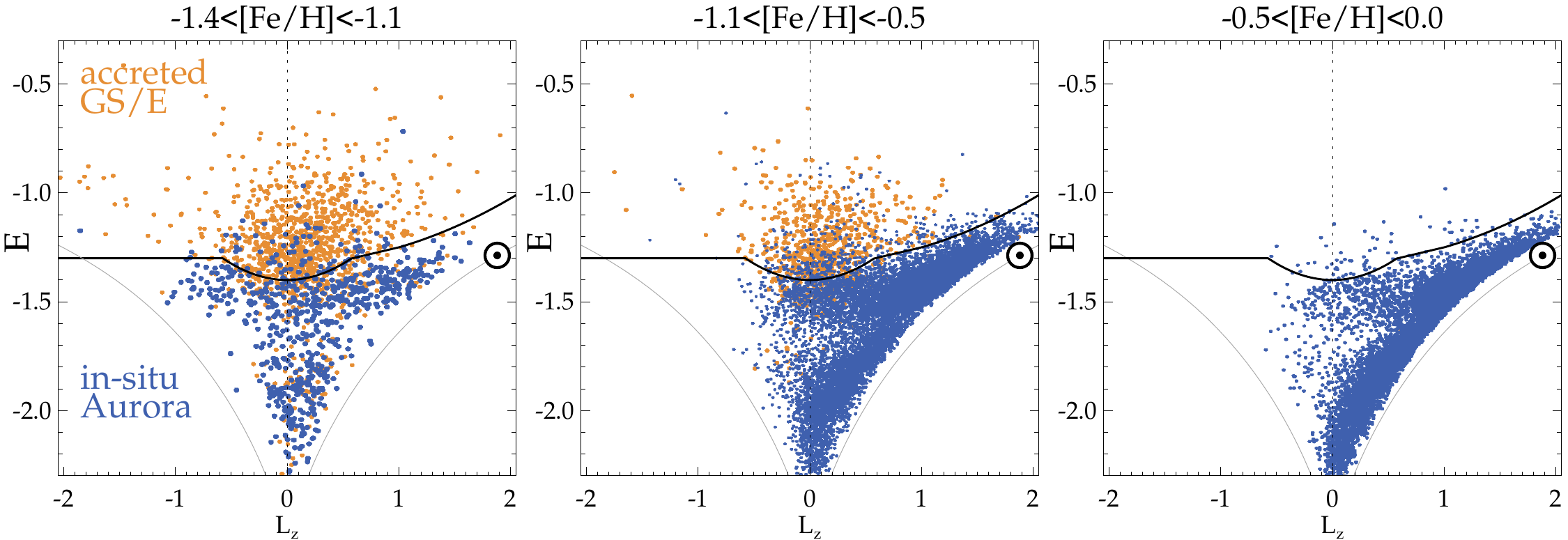}
  \caption[]{Orbital properties of halo giants. {\bf Left:} Distribution of Aurora (blue, selected with $\rm [Al/Fe]>-0.075$) and accreted (orange, selected with $\rm [Al/Fe]<-0.075$ and an additional [Mg/Fe] cut - see text for details) giants with $-1.4<\rm [Fe/H]<-1.1$ in the space of energy $E$ ($\times10^{5}$) and vertical component of angular momentum $L_z$ ($\times10^{3}$).  Solid black line marks the decision boundary to separate stars into accreted (high energy) and Aurora (in-situ, low energy populations. Grey solid lines correspond to the maximal angular momentum at fixed energy, i.e. orbits with circular velocity $V_{\phi}=V_{\rm circ}$. $\odot$ marks the location of the Sun in the chosen potential. {\bf Middle:} Same as Left but for stars with $-1.1<\rm [Fe/H]<-0.5$. {\bf Right:} Same as previous panels but for $-0.5<\rm [Fe/H]<0$, note that in this metallicity range no accreted stars are visible.}
   \label{fig:elz_aurora}
\end{figure*}

As demonstrated by panel c of the Figure, the number of high-[N/O] GC-like stars varies significantly with metallicity [Fe/H]. Black lines give the 20th, 50th and 80th percentiles of the [N/O] distribution as a function of [Fe/H]. At higher metallicities, i.e. at [Fe/H]$>-0.5$, the average [N/O] level starts to climb up due to the increased nitrogen contribution from intermediate-mass Asymptotic Giant Branch (AGB) stars \citep[see, e.g.,][]{Kobayashi2020,Johnson2023}. These nitrogen-rich and metal-rich stars are clearly distinct from the GC-born high-[N/O] stars because they do not exhibit any strong (anti)correlations typical of clusters (e.g., between Mg and Al). The bulk of these high-metallicity AGB-contributed disk stars with elevated [N/O] are removed by our $V_{\rm \phi}<160$ km/s cut. Also note a slight upward [N/O] trend with decreasing [Fe/H]. At least in part this is caused by culling measurements with high uncertainties. This weak [N/O] trend is however much flatter than a more noticeable increase in median [N/Fe] at low [Fe/H] as shown in panel d of the Figure \citep[see also Figure 1 of][]{Schiavon2017}.

We proceed by classifying the field high-[N/O] GC-like stars into those born in-situ in the Milky Way (including Aurora) and those formed in dwarf galaxies  (mainly GS/E) and subsequently incorporated into the accreted stellar halo.

\subsection{Distinguishing accreted and in-situ  stars and clusters}

Due to a relatively slow pace of star formation and as a result of a strong metallicity dependence of the Al yield, dwarf galaxies never experience an over-abundance of Al compared to Fe, with the ratio [Al/Fe] staying low across a wide range of [Fe/H]. On the contrary, stars born in the Milky Way exhibit a rapid increase in [Al/Fe] around $-1.5\rm <[Fe/H]<-0.9$. As shown in \citet{Hawkins2015}, distinct behaviour of [Al/Fe] (and [Na/Fe]) can be utilized to separate accreted and in-situ halo components \citep[see also][]{Das2020}. 

Specifically, BK22 classify stars with $\rm [Al/Fe]>-0.075$ as in-situ and those with $\rm [Al/Fe]<-0.075$ as accreted. This approximate classification is supported by the observed abundance trends in the surviving massive MW dwarf satellites that typically have $\rm [Al/Fe]<-0.1$, as reported by \citet{Hasselquist2021}. At low metallicities, $\rm [Fe/H]\lesssim-1.5$, the use of [Al/Fe] is not viable as the in-situ and the accreted sequences start to merge. More importantly, [Al/Fe]-based classification into accreted and in-situ is not viable for GC-like high-[N/O] stars due to their anomalously high [Al/Fe] ratios, $\rm [Al/Fe]>-0.1$.

We thus adopt a different two-stage approach to classify stars and GCs into in-situ and accreted populations. We first determine a boundary in the $E-L_z$ space that separates in-situ and accreted objects for the stars without anomalous ratios and at metallicities where [Al/Fe]-based classification is robust ($-1.4<{\rm [Fe/H]}<-1.1$), as shown in the left panel of  
Figure~\ref{fig:elz_aurora}. The panel shows $E$ and $L_z$ distributions of stars classified as accreted (primarily GS/E, orange) and in-situ (Aurora, blue) using the $\rm [Al/Fe]=-0.075$ threshold. In addition, following BK22  we apply the cuts of  $\rm [Mg/Fe]<-0.3\,[Fe/H]-0.1$ to the accreted stars. 

As the left panel of Figure~\ref{fig:elz_aurora} reveals,  the distribution of accreted (GS/E) stars has a narrow range of $z-$component of angular momenta $\vert L_z\vert<0.6$ and is limited in energy to $E\gtrsim -1.4$. This is in agreement with \citet{wrinkles}, where a stellar sample based on the {\it Gaia} DR3 data was used for the analysis. The in-situ stars (blue points) on the other hand have a broader $L_z$ distribution at energies similar to the lowest levels reached by the GS/E stars, i.e. $E\sim-1.4$ where the two groups have a small overlap. In-situ stars continue to lower energies where the occurrence of accreted stars (classified with [Al/Fe] cut) is negligible and is likely to an occasional scatter of [Al/Fe] values below the threshold. 

\begin{figure*}
  \centering
  \includegraphics[width=0.99\textwidth]{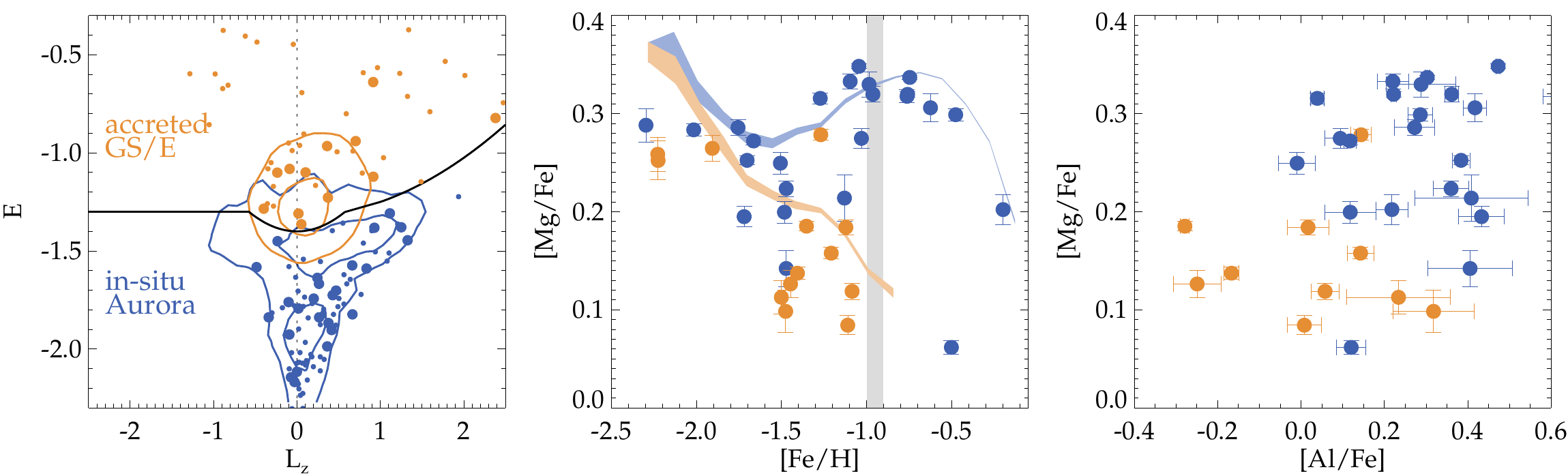}
  \caption[]{Properties of globular clusters classified as accreted (orange) and in-situ (including Aurora, blue). {\bf Left:} Distribution of all Galactic globular clusters (small points) in the space of $E$ and $L_z$ (similar to left panel of Figure~\ref{fig:elz_aurora}. Accreted (in-situ, including Aurora) GCs with high-quality $f_{\rm N/O}$ measurements are shown as orange (blue) filled circles. Blue (orange) contours show density of in-situ (accreted) stars from the left panel of Figure~\ref{fig:elz_aurora}. {\bf Middle:} Median [Mg/Fe] abundances with associated uncertainties as a function of cluster's [Fe/H] for accreted and Aurora GCs, classified as shown in the Left panel. Only GCs with median Mg uncertainties less than 0.025 dex are shown. Note that GCs classified as in-situ have systematically higher [Mg/Fe] compared to those classified as accreted. Orange (blue) bands show median stellar [Mg/Fe] abundance ratios (with associated uncertainties) for field stars as a function of metallicity. These stars have been classified using the same $E, L_z$ boundary shown in the left panel. Note a clear peak in [Mg/Fe] of both in-situ GCs and in-situ field stars around [Fe/H]$\approx-1$ (the Spin-up, marked by a grey vertical band). {\bf Right:} Orange (blue) points show median [Mg/Fe] as a function of median [Al/Fe] for stars in the Galactic GCs observed by APOGEE and classified as accreted (in-situ).}
   \label{fig:gc_aur_acc}
\end{figure*}

The solid line approximates 
the boundary separating the in-situ and accreted populations in the $E-L_z$ space visible in the left panel and is described by the following equation: 
\begin{align}
\label{eq:sel}
      L_z<-0.58:~ &	E=-1.3 \nonumber \\
-0.58<L_z<0.58:~ &	E=-1.4+0.3L_z^2\\
          L_z>0.58:~	&	E=-1.325+0.075L_z^2, \nonumber
\end{align}
where $E$ is in units of $10^5\, \mathrm{km}^2\,\mathrm{s}^{-2}$ and $L_z$ is in units of $10^3\,\mathrm{kpc}\,\mathrm{km}\mathrm{ s}^{-1}$. 
We then test that the same boundary separates in-situ and accreted populations at other metallicities. 

Middle panel of Figure~\ref{fig:elz_aurora} considers high metallicity stars, i.e. $-1.1<\rm [Fe/H]<-0.5$. While some GS/E debris is still visible, the $E$, $L_z$ distribution is dominated by the in-situ stars, mainly high-$\alpha$ disk and the Splash \citep[see][]{Splash}. Finally, the right panel of the Figure gives the distribution of stars with $\rm [Fe/H]>-0.5$ and, unsurprisingly, shows no presence of any accreted debris. 
As we can see the same boundary of equation~\ref{eq:sel}, shown by the solid line, separates the in-situ and accreted components well. 
Thus, in what follows, we use this boundary to classify both stars and GCs of all metallicities into in-situ and accreted.

We note that the dominance of high-$\alpha$ and low-[Fe/H] Aurora at low energies explains the trends reported in \citet{Donlon2023} without invoking additional accretion events.

Although the boundary in Equation~\ref{eq:sel} is categorical and was derived as a simple approximation to the distribution of the in-situ and accreted stars, we have also tested it using
machine learning classification. Specifically, we used the ``Gradient Boosted Trees'' (GBT) machine learning method  \citep{Friedman.2001}, implemented in the {\tt GradientBoostingClassifier} class in the Sci-kit Learn package \citep{sklearn}, which allows for overlap between classes by assigning a class probability for stars in the overlap region. The method was trained with 80\% of the sample of stars with reliable [Al/Fe]-based classification, while the remaining 20\% of stars was used to test classification accuracy. We then computed the classification accuracy obtained with GBT  and with the categorical boundary of Equation~\ref{eq:sel} for the test sample finding that both methods result in $\approx 96\%$ accuracy in classification. We also tested other machine learning methods, such as Extremely Random Trees and artificial neural networks finding comparable or lower accuracy. 
Thus, the accuracy of classification with Equation~\ref{eq:sel} is comparable with the accuracy of classification with machine learning methods.

\subsection{Distinct chemical properties of accreted and in-situ GCs}

Figure~\ref{fig:gc_aur_acc} tests our accreted/in-situ classification based on the position in the $E, L_z$ space on the Galactic GCs. Left panel of the Figure shows positions of all GCs with measured orbital properties as small coloured dots. Orange marks the accreted GCs and blue the in-situ ones (including Aurora).  Out of 149 Galactic GCs considered, 98, or $\approx 2/3$ are classified as in-situ. GCs with sufficient APOGEE measurements are shown as large filled circles coloured according to their classification. There are 25 in-situ and 13 accreted GCs in our APOGEE sample. For further analysis we only consider GCs with sufficient number of measurements. For example, at least 3 GC member stars are required when studying GC properties reported in this Section; when the high-[N/O] stars are concerned, at least 10 candidate members are required as well as the relative uncertainty on the fraction of the high-[N/O] stars $f_{\rm N/O}$ (selected according to the conditions stipulated in Section~\ref{sec:no_select}), less than 50\%. The latter combination of cuts leaves only 28 clusters, of which 11 are classified as accreted and 17 as in-situ.

The middle panel of Figure~\ref{fig:gc_aur_acc} shows the median [Mg/Fe] ratios (with associated uncertainties) for the accreted and in-situ GCs as a function of metallicity. Note that in this Figure we show the median metallicity of the cluster's APOGEE member stars. There is little overlap between the two groups. The accreted GCs typically have median [Mg/Fe] ratios lower by some 0.15 dex than those of the in-situ clusters. The [Mg/Fe] values in Galactic GCs can be compared to the halo stars across the same metallicity range separated into accreted (orange band) and in-situ (blue band) using the same $E, L_z$ criteria. Reassuringly, halo stars follow the same trends in the space of [Mg/Fe] and [Fe/H], in particular, at [Fe/H]$>-1.5$, median [Mg/Fe] for in-situ stars is $\sim0.15$ dex higher than that for the accreted population \citep[a similar conclusion is reached in][]{Horta_GC}. 

There is a striking trend of [Mg/Fe] with increasing metallicity exhibited by both field in-situ stars and the MW in-situ GCs in the middle panel of Figure~\ref{fig:gc_aur_acc}. At $\rm [Fe/H]\approx-1.3$, [Mg/Fe] starts to rise and exhibits a broad peak at $\rm [Fe/H]\approx-1\div -0.7$. The magnitude of this [Mg/Fe] increase is modest, just under 0.1 dex. Beyond $\rm [Fe/H]\approx -0.7$, the [Mg/Fe] ratio is decreasing. As traced by field stars, this [Mg/Fe] peak can also be seen in Figure 7 of BK22 (see also the left panel of Figure~\ref{fig:no_feh_elz} below) and is discussed at length in \citet{Conroy2022}. Here for the first time, we show that the same pattern is displayed by the Galactic in-situ GCs. 

\begin{figure*}
  \centering
  \includegraphics[width=0.99\textwidth]{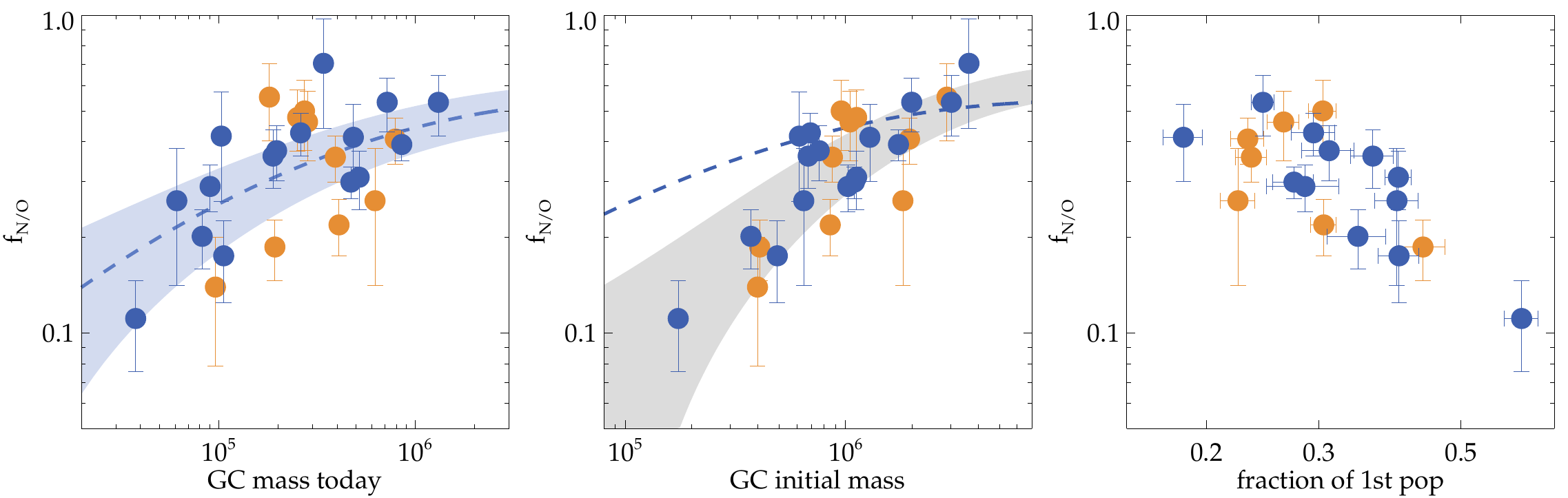}
  \caption[]{Fraction of high-[N/O] stars $f_{\rm N/O}$ in the Milky Way globular clusters. {\bf Left:} $f_{\rm N/O}$ as a function of the present-day GC mass. Blue dashed line and band show quadratic fit to measurements of the in-situ GCs. {\bf Middle:} $f_{\rm N/O}$ as a function of the GC's initial mass, as estimated by \citet{Baumgardt2003}. The dashed blue line shows the quadratic fit to the correlation with present-day mass in the left panel. The grey band shows an approximate quadratic fit to all GCs shown in this panel. {\bf Right:} $f_{\rm N/O}$ as a function of the fraction of the GC's 1st population stars, as measured by \citet{Milone2017}. }
   \label{fig:gc_mass}
\end{figure*}

As elucidated in \citet{Weinberg2017}, bumps in $\alpha$-element ratios of the order of $0.1-0.3$ dex are a tell-tale sign of a star-formation burst in a system converting a sizeable portion of its gas into stars, whilst retaining the core-collapse supernova enrichment products. $\alpha$-bumps similar to that reported here for the Galactic in-situ stars have been seen in several massive MW dwarf satellites by \citet{Hasselquist2021}. \citet{Conroy2022} suggest that the MW in-situ $\alpha$-bump can be explained by models in which the star formation efficiency increases sharply by a factor of $\sim$10 around $\rm [Fe/H]\approx-1$. As the middle panel of Figure~\ref{fig:gc_aur_acc} demonstrates, below $\rm [Fe/H]=-1.8$, both the in-situ and the accreted stars show a systematic and noticeable increase in [Mg/Fe]. We believe that this [Mg/Fe] rise at low [Fe/H] may not be a genuine characteristic of the Galactic field stars but rather a sign of the APOGEE pipeline struggling with measurements of low and intermediate $\alpha$ abundances at low [Fe/H]. This hypothesis is based on the following tests: i) the media GC values show a less pronounced rise compared to the field stars, ii) the increases at low metallicities is reduced if no cuts are applied on abundance uncertainties or other effective signal-to-noise filters, and iii) [Si/Fe] shows a much flatter behaviour at [Fe/H]$<-1.5$.

Right panel of Figure~\ref{fig:gc_aur_acc} gives the distribution of the in-situ (blue) and accreted (orange) GCs in the space of median [Al/Fe] and [Mg/Fe]. GCs with [Fe/H]$<-1.8$ are excluded due to the suspected bias in [Mg/Fe] measurements mentioned above. 
The in-situ and accreted GC populations occupy distinct regions of [Mg/Fe]-[Al/Fe] space and show little overlap. On average, the in-situ GC have higher values of [Al/Fe]. At fixed [Al/Fe], the in-situ GCs have higher [Mg/Fe] ratios. The only two objects that buck this trend are NGC 6388 (classified as in-situ, blue point at $\rm [Mg/Fe]\approx0$) and NGC 288 (classified as accreted, orange point with $\rm [Mg/Fe]>0.25$). The peculiar properties of these two clusters have been noted before and are discussed in the literature \citep[see e.g.][]{Myeong2019, Massari2019, Horta_GC}. Most recently, \citet{Carretta2022} argued for the in-situ origin of NGC 6388 based on the abundance pattern of iron-peak elements. For NGC 288, \citet{Monty2023} show that it does not follow the chemical trends of other GS/E GCs.

As revealed by GCs' detailed chemical properties, the misclassification rate is at a level similar to that indicated by our experiments with the GBT ML method. Additionally, we note that including a rotating bar can and will affect some of the GCs' orbits and consequently might change their membership in the accreted/in-situ groups. However, when the effect of the bar is included, only 3 out of 40 clusters considered in the study of \citet{Perez2020} have been singled out as potential outer halo interlopers. 

Notwithstanding these intricacies, the middle and right panels of Figure~\ref{fig:gc_aur_acc} show that the simple $E,L_z$-based classification of stars and GCs into accreted (primarily GS/E) and in-situ (primarily Aurora) using boundary defined by equation~\ref{eq:sel} works well. In particular, this classification results in populations of GCs with distinct chemical properties:  
the in-situ GCs exhibit systematically higher Mg abundance and the two GC classes occupy distinct regions in the $\rm [Al/Fe]-[Mg/Fe]$ space. 

\subsection{Trends of high-N fraction with cluster properties}
\label{sec:trends}

Figure~\ref{fig:gc_mass} explores the behaviour of the fraction of high-[N/O] stars in the Galactic GCs observed by APOGEE DR17. We have excluded NGC 6715 (M 54) residing in the core of the Sgr dwarf galaxy, as well as NGC 5139 ($\omega$ Cen). The left panel of the Figure shows $f_{\rm N/O}$ as a function of the cluster's present-day mass. While there is a considerable scatter, there is a clear trend of $f_{\rm N/O}$ increasing with increasing cluster mass, from $\approx 10\%$ in  $M\approx10^5 M_{\odot}$ clusters to $f_{\rm N/O}\approx50\%$ in clusters of $M\approx10^6 M_{\odot}$. There is also an indication that the trend is more pronounced for the in-situ GCs (shown in blue). 

\begin{figure*}
  \centering
  \includegraphics[width=0.99\textwidth]{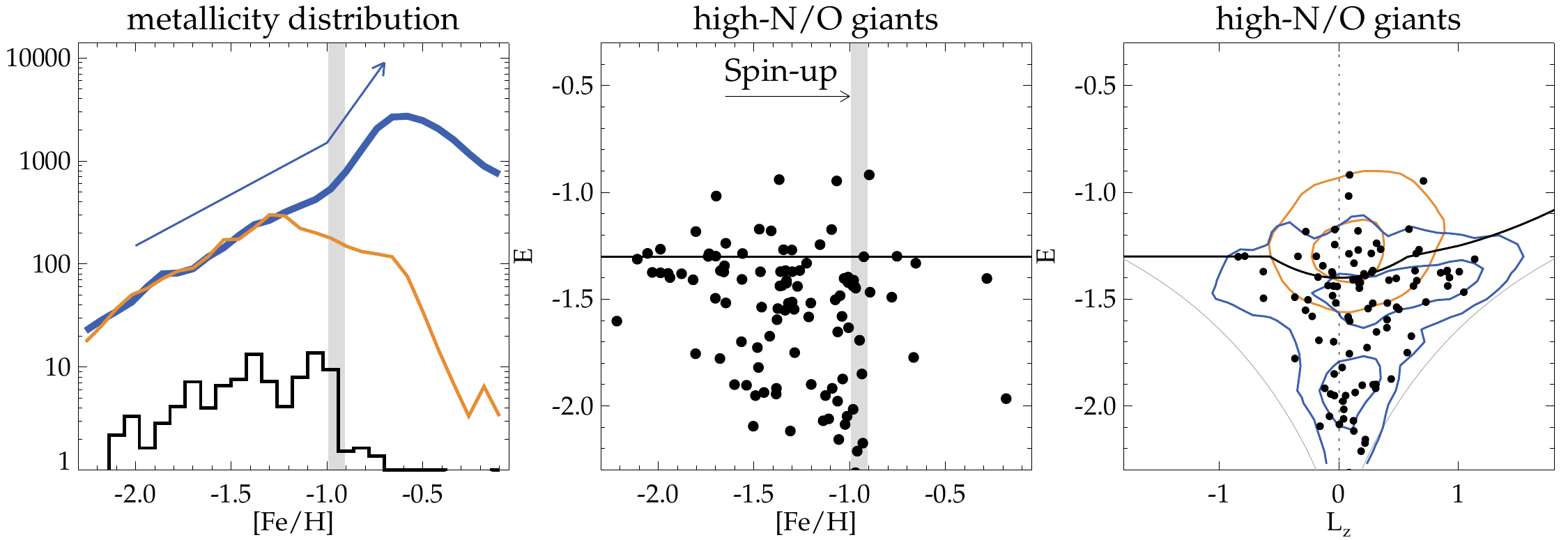}
  \caption[]{{\bf Left:} Metallicity distribution of high-[N/O] field giants (black) is compared to the metallicity distribution of field giants classified as accreted (orange) and in-situ (blue). Only stars with low azimuthal velocities are used. Around $\rm [Fe/H]\approx-1$ (Spin-up, marked by the vertical grey band) there is a sharp upward bend in the slope of the distribution of the in-situ field stars (as indicated by the blue arrow) and a sharp truncation in the metallicity distribution of the high-[N/O] stars. {\bf Middle:} Energy of high-[N/O] stars as a function of metallicity. The horizontal black line shows the approximate energy threshold below which the majority of stars are of in-situ nature. {\bf Right:} $E, Lz$ distribution of high-[N/O] stars (black dots). Blue (orange) contours show the density of in-situ (accreted) stars from the left panel of Figure~\ref{fig:elz_aurora}.}
   \label{fig:no_feh_elz}
\end{figure*}
\begin{figure}
  \centering
  \includegraphics[width=0.49\textwidth]{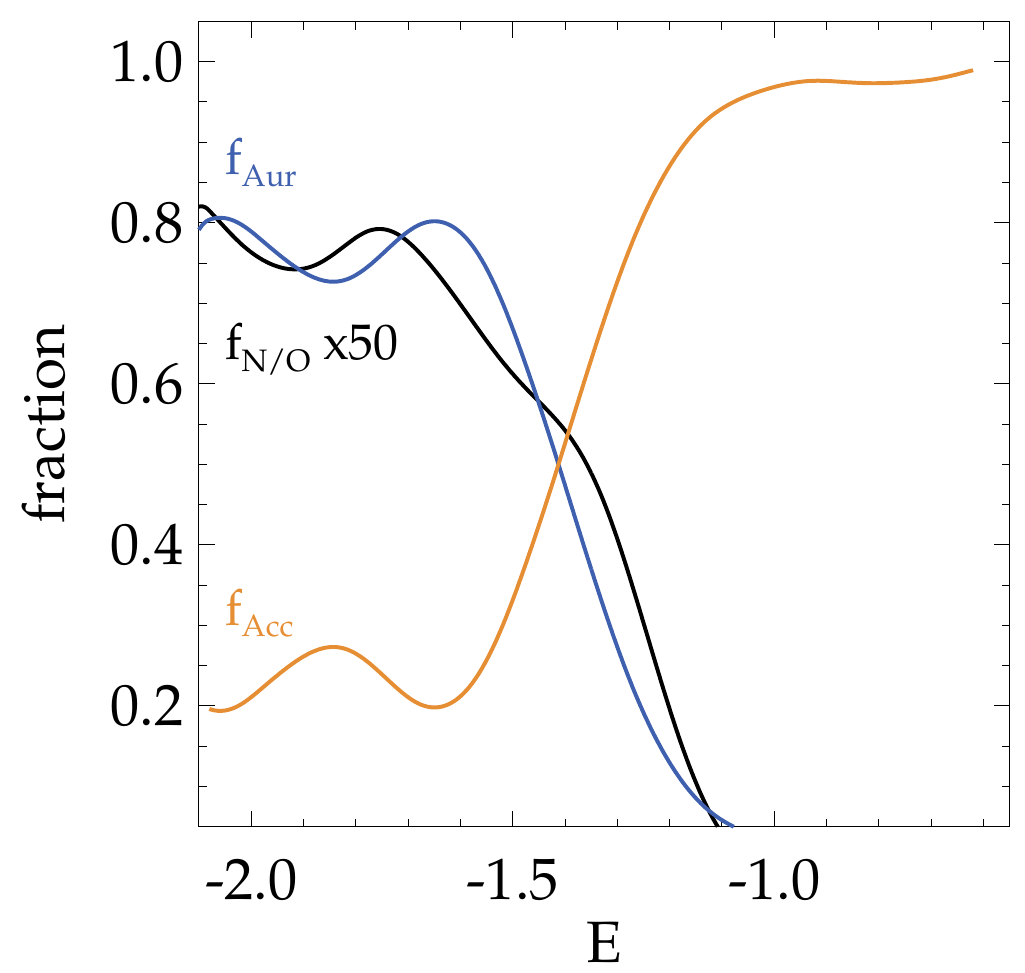}
  \caption[]{Fraction of Aurora (accreted) stars shown in blue (orange) and fraction of high-[N/O] GC-like giants (multiplied by 50) shown in black as a function of energy. The accreted, in-situ and high-[N/O] samples are the same as shown in Figure~\ref{fig:elz_aurora}. Note the similarity of  $f_{\rm N/O}$ and $f_{\rm Aurora}$ dependencies on energy. For comparison, dashed grey line gives the energy distribution of high-[N/O] stars with low-[Al/Fe], i.e. those likely not born in GCs.}
   \label{fig:frac_energy}
\end{figure}

A second-degree polynomial fit ($\pm0.075$) to the in-situ $f_{\rm N/O}$ values is shown as a blue band. The middle panel of the Figure shows that there is also a trend of increasing $f_{\rm N/O}$ with the GC's initial mass. The dependence on the initial mass appears tighter and stronger compared to the correlation with the present-day mass shown in the left panel. The quadratic fit to all of the available data is shown as a grey band which includes $\pm0.075$ scatter around the mean model. 

Finally, the right panel of Figure~\ref{fig:gc_mass} shows that $f_{\rm N/O}$ decreases with decreasing fraction of the 1st stellar population in clusters \citep[where available, as measured by][]{Milone2017}. There is no obvious difference between in-situ and accreted GCs in terms of the trends of $f_{\rm N/O}$ as a function of GC mass or fraction of 1st population in agreement with results of \citet{Milone2020}.  

The trends shown in Figure~\ref{fig:gc_mass}  are consistent with those previously reported in the literature \citep[see e.g.][]{Bastian_Lardo2018,Gratton2019}. Note however that the above dependence on the initial GC mass allows us to calibrate the $f_{\rm N/O}$ (computed, as many previous studies, according to a somewhat arbitrary threshold) and thus link it directly to the Galaxy's original cluster population.

\subsection{The majority of high-[N/O] field stars belong to Aurora}

Left panel of Figure~\ref{fig:no_feh_elz} shows metallicity distributions of  in-situ (blue), accreted (orage), and high-[N/O] stars (black histogram). The high-[N/O] sample shows a sharp truncation at $\rm [Fe/H]\approx-1$, a broad peak at $-1.5\lesssim\rm Fe/H]\lesssim-1$ and a considerable drop off towards $\rm [Fe/H]\approx-2$. All populations, i.e. in-situ, accreted and high-[N/O] stars have approximately the same slope of the metallicity distribution at $\rm [Fe/H]<-1$ \citep[see discussion in][]{Rix2022}. At $\rm[Fe/H]\approx-1$, the in-situ distribution shows a sharp upward inflection matching the location of the cliff in the [Fe/H] distribution of the high-[N/O] stars \citep[see also Figure 4 in][]{Aurora}. At $\rm[Fe/H]\approx-1$, the slope of the in-situ metallicity distribution changes abruptly from ${\rm d} \log(n)/{\rm d[Fe/H]}\approx1$ to ${\rm d} \log(n)/{\rm d[Fe/H]}\approx2.3$. The metallicity distribution of the accreted stars does not show any obvious features around $\rm [Fe/H]\approx -1$, instead it exhibits a continuous downward slope from $\rm [Fe/H]\approx-1.3$ to $\rm [Fe/H]\approx-0.7$ where it is sharply truncated \citep[in agreement with other studies, see, e.g,][]{Belokurov2018,Mackereth2019,Splash,Feuillet2020,Sanders2021}.

The middle panel of Figure~\ref{fig:no_feh_elz} shows the distribution of the high-[N/O] stars in the $E-\rm [Fe/H]$ space. Apart from the sharp truncation in the number of high-[N/O] stars at $\rm [Fe/H]\approx-1$ at all energies, there does not seem to be any strong correlation between $E$ and [Fe/H]. Note, however, that at $\rm [Fe/H]<-1.7$ the selected high-[N/O] stars do not span the entire range of energies of the high-metallicity counterparts. It is difficult to assess whether this is a sign of a genuine change in the properties of the high-[N/O] population or is simply a result of small statistics of stars in our sample.

The right panel of Figure~\ref{fig:no_feh_elz} shows the $E-L_z$ distribution of the GC-like high-[N/O] stars selected as described in Section~\ref{sec:no_select}. The absolute majority of the stars fall within the blue contours marking the distribution of the in-situ stars.
This indicates that most of high-[N/O] stars belong to the Aurora population of low-metallicity in-situ stars. 

This is further clarified in Figure~\ref{fig:frac_energy}, which shows the fraction of the Aurora, accreted, and high-[N/O] stars as a function of their total energy $E$. Note that at low energies a few stars designated as accreted are likely misclassified in-situ stars. Due to the sharp rise in [Al/Fe] with increasing [Fe/H], some of the low-metallicity Aurora stars drop below the nominal threshold.  
In the intermediate range of $-1.4<E<-1.3$, the typical contribution of Aurora stars is $\approx 45\%$. The fraction of high-[N/O] stars is always low, typically $\sim1\%$. Figure~\ref{fig:frac_energy} shows that in the metallicity range considered $\approx 90\%$ of stars with $E>-1.2$ belong to the accreted population. 

The striking fact shown in Figure~\ref{fig:frac_energy} is that trends of $f_{\rm N/O}$ and $f_{\rm Aurora}$ with energy are remarkably similar. As the energy decreases, the fraction of in-situ stars grows sharply, reaching $\approx 80\%$ at $E\approx-1.5$. This is a clear indication that these stellar populations are closely related. 

\subsection{Fraction of field high-[N/O] stars as a function of [Fe/H]}
\label{sec:frac_feh}

\begin{figure}
  \centering
  \includegraphics[width=0.49\textwidth]{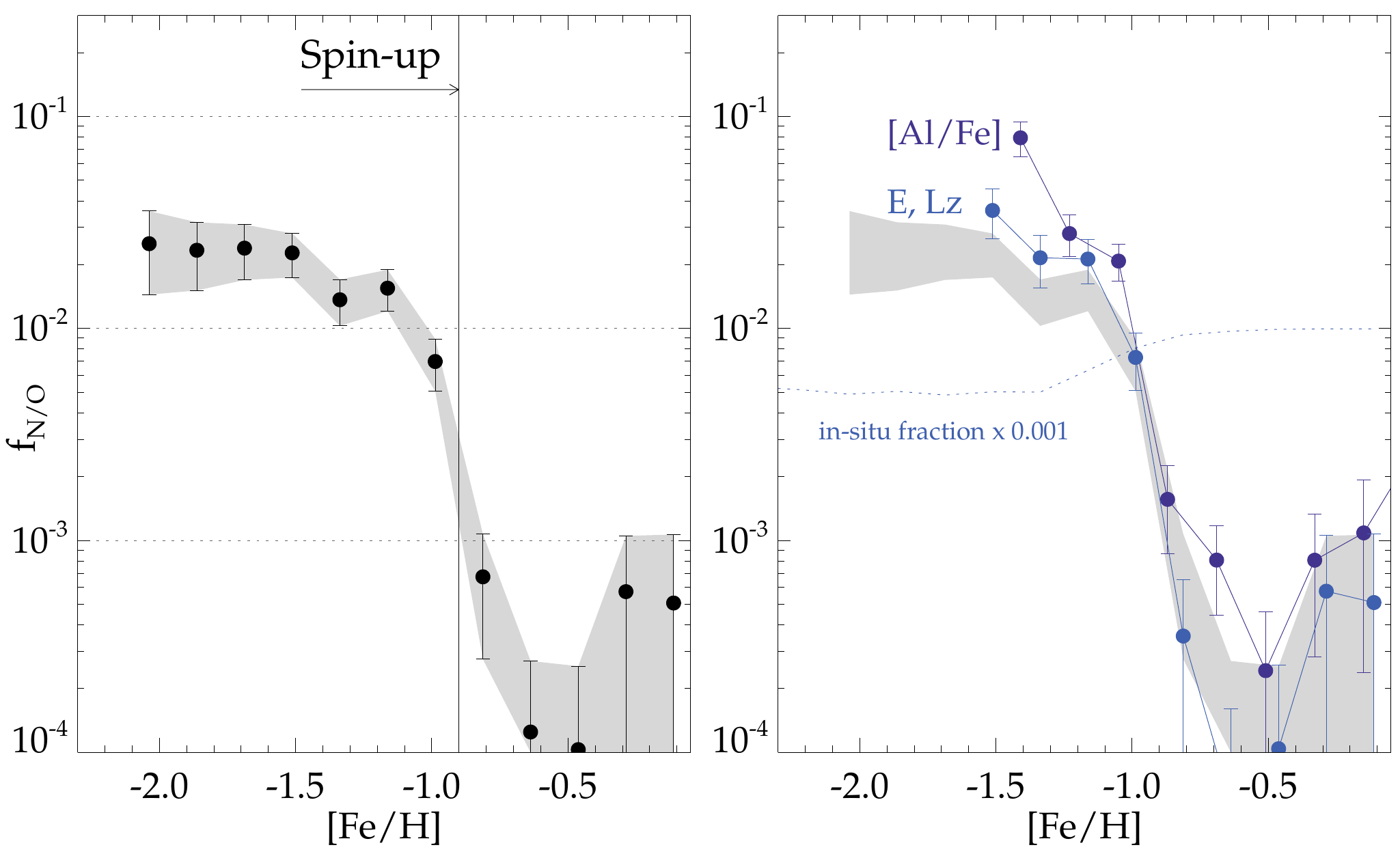}
  \caption[]{{\bf Left:} Fraction of high-[N/O] stars $f_{\rm N/O}$ with low tangential velocities as a function of metallicity. Stars in globular clusters and Magellanic Clouds have been excluded. Note the abrupt jump in $f_{\rm N/O}$ around $\rm [Fe/H]\approx-0.9$. {\bf Right:} Comparison of the in-situ $f_{\rm N/O}$ ratio computed with two different methods of in-situ classification. Blue (purple) shows classification based on $(E,L_z)$ (ratio of [Al/Fe]). The two estimates agree for $\rm [Fe/H]>-1.4$. Around $\rm [Fe/H]\sim-1.4$ the [Al/Fe] ratio of the Aurora population drops quickly under the nominal threshold used to separate the accreted and the in-situ stars. As a result, the $f_{\rm N/O}$ ratio artificially increases. Blue dotted line shows the re-scaled in-situ fraction, note the drop from 100\% to 50\% at $-1.5<$[Fe/H]$<-1$.}
   \label{fig:frac_feh}
\end{figure}

Figure~\ref{fig:frac_feh} shows  the fraction of high-[N/O] field stars in the Milky Way as a function of metallicity. At $\rm [Fe/H]<-1$, the fraction is flat at approximately $1\%<f_{\rm N/O}<2\%$. In contrast, at $\rm[Fe/H]\sim-0.9$, $f_{\rm N/O}$ drops abruptly, decreasing with increasing [Fe/H] by an order of magnitude. This dramatic change in the $f_{\rm N/O}$ with metallicity mirrors chemo-kinematic trends observed in the in-situ stars in BK22. They associate a rapid change in the chemical abundance spreads and the overall spin of the Galaxy's stars with the Milky Way transitioning from a chaotic pre-disk state into a coherently rotating disk. 

\begin{figure}
  \centering
  \includegraphics[width=0.49\textwidth]{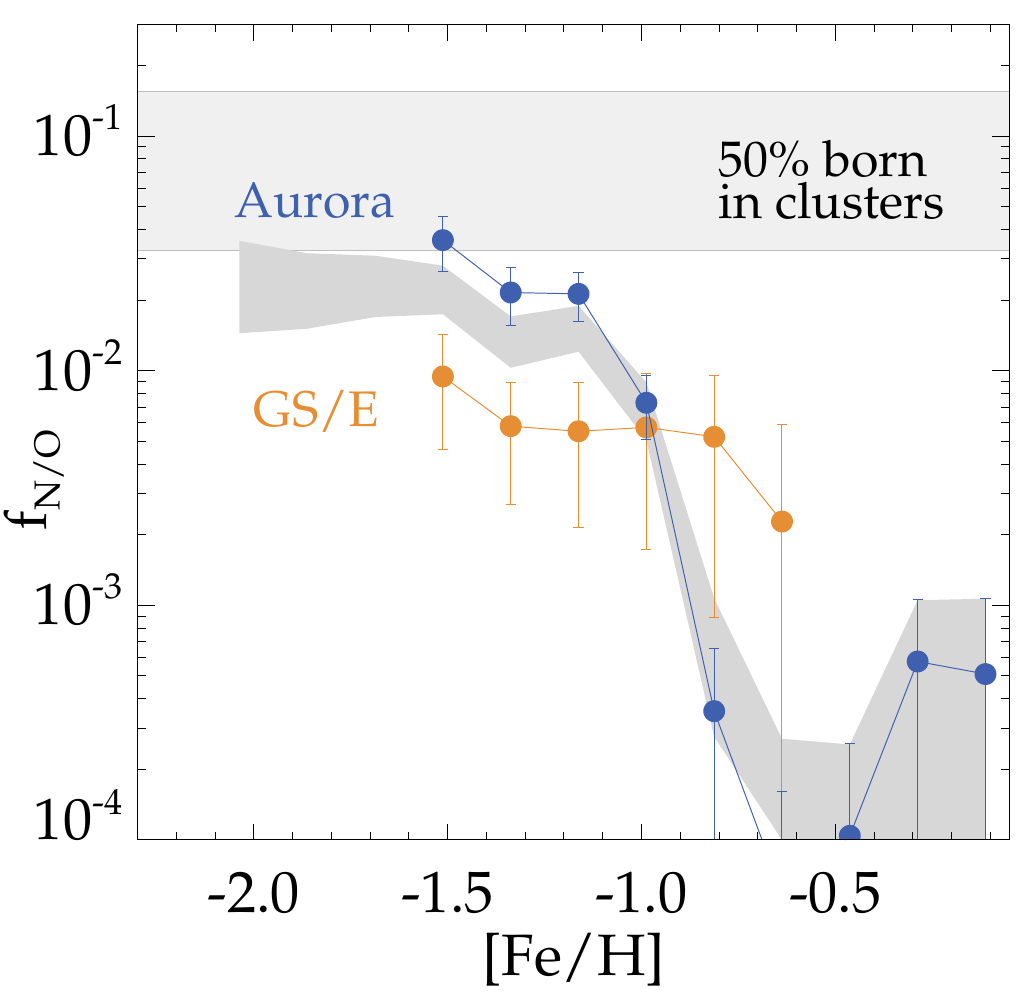}
  \caption[]{Grey band shows $f_{\rm N/O}$ measurements reported in Figure~\ref{fig:frac_feh}. Blue (orange) line and symbols show measurements for in-situ (accreted) populations classified using energy boundary shown in Figure~\ref{fig:elz_aurora}. Below [Fe/H]=-0.9, Aurora's $f_{\rm N/O}$ is $>3$ times higher than that for the accreted stars. Above [Fe/H]=-0.9, Aurora's $f_{\rm N/O}$ plummets by more than an order of magnitude, while the behaviour of the accreted population population is much flatter. At [Fe/H]$<-1$, Aurora's $f_{\rm N/O}=0.04$, which is just under a half of the typical fraction in our fiducial model (see text for details), implying that close to a half of the Aurora stars (at this metallicity) were born in clusters. }
   \label{fig:frac_feh2}
\end{figure}

Indeed, Figure~\ref{fig:frac_feh2} shows that the sharp increase in $f_{\rm N/O}$ is due entirely to the stars in the Aurora population, while $f_{\rm N/O}$ of the accreted (primarily GS/E) population does not change significantly with metallicity. Moreover, the in-situ population (Aurora) shows $\approx 5$ times higher fraction of high-[N/O] stars compared to that of the GS/E. This increase somewhat depends on the selection method of the Aurora population (using $E,L_z$ or [Al/Fe]), as shown in the right panel of Figure~\ref{fig:frac_feh}.  Reassuringly, the two methods ($E, L_z$-based and [Al/Fe]-based) yield $f_{\rm N/O}$ curves that are similar across a wide range of metallicity, i.e for $\rm [Fe/H]>-1.3$. Around $\rm [Fe/H]\approx-1.4$ the [Al/Fe] ratio of the in-situ stars starts to drop below the chosen threshold (see Figure 2 in BK22) and, as a consequence, the [Al/Fe]-based in-situ $f_{\rm N/O}$ is biased high compared to the $E,L_z$ one at this metallicity. However, these differences are much smaller than the overall trend itself and can be viewed as uncertainty with which $f_{\rm N/O}$ is estimated for the Aurora population. 

The sharp increase of the fraction of high-[N/O] Aurora stars $f_{\rm N/O}$ with decreasing [Fe/H] shows that processes that produced anomalous amounts of nitrogen in the regions where high-[N/O] stars were born were much more prevalent at lower metallicities. If massive star clusters are plausible sites of high-[N/O] star formation, we can expect that the metallicity distribution of the high-[N/O] stars is similar to that of the globular clusters. 

In principle, if high-[N/O] were produced in clusters, their metallicity distribution should trace distribution of metallicity of the disrupted clusters. Moreover, given that $f_{\rm N/O}$ is higher in larger mass clusters (Figure~\ref{fig:gc_mass}) and if the fraction of massive clusters had a dependence on metallicity, the metallicity distribution of field high-[N/O] stars could be somewhat different from the metallicity distribution of clusters. However, models indicate that the fraction of disrupted clusters is nearly independent of metallicity (O. Gnedin, priv. communication) and there is no significant trend of cluster mass distribution with metallicity for observed surviving GCs. Thus, we can generally expect the metallicity distribution of high-[N/O] stars and {\it surviving} clusters to be quite similar. 

\begin{figure*}
  \centering
  \includegraphics[width=\textwidth]{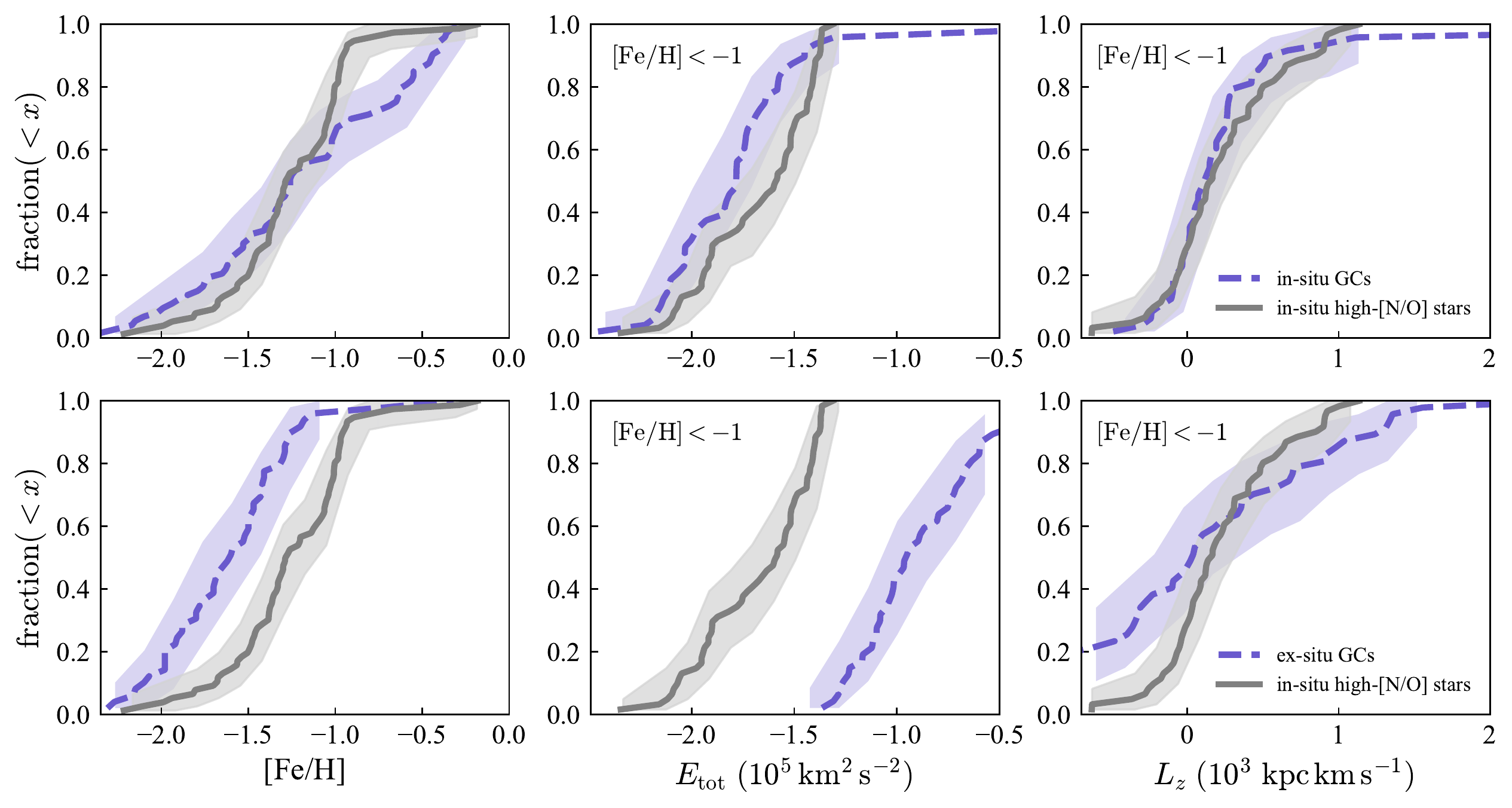}
  \caption[]{Cumulative distributions of metallicity ([Fe/H]), energy and $L_z$ component of the angular momentum of the  high-[N/O] stars (gray solid lines and shaded bands) and MW globular clusters (blue dashed lines and shaded bands). To account 
  for the velocity cut $V_\phi<160$ km/s for the stars, we applied the same cut to GCs for this comparison. The gray and blue shaded bands show corresponding 95\% range estimated using bootstrap resampling. The top (bottom) row compares distributions of the 
  in-situ  high-[N/O] stars with in-situ (ex-situ) GCs. Note that [Fe/H] distributions are compared for the full range of metallicity, while $E_{\rm tot}$ and $L_z$ distributions are compared for stars and GCs in the metallicity range where metallicity distributions of in-situ stars and GCs agree (upper left panel): $\rm [Fe/H]<-1$. 
 The figure shows that distributions of metallicity, energy, and momentum of the in-situ GCs and in-situ high-[N/O] stars are similar in this metallicity range, which indicates that these stars originated from low-metallicity GCs formed in the Milky Way. }
   \label{fig:feh_dist_ns_gc}
\end{figure*}

Indeed, the upper right panel in Figure~\ref{fig:feh_dist_ns_gc} shows that cumulative metallicity distributions of the in-situ high-[N/O] stars and in-situ MW's globular clusters are quite similar for stars and GCs with $\rm [Fe/H]<-1$ and differ at larger metallicities. For this comparison we applied a cut of $V_{\phi}$ to the sample of GCs to account for a similar cut used in selecting high-[N/O] stats. The similarity of metallicity distributions is confirmed by the Kolmogorov-Smirnov (KS) probability that the two distribution agree at $\rm [Fe/H]<-1$. The 95\% range of this probability evaluated using bootstrap resampling is extending to probability of $0.21$. In contrast, the metallicity distribution of the in-situ high-[N/O] stars is different from that of the ex-situ GCs and the 95\% range of the KS probability extends only to 0.01 (with the mean value of $\approx 10^{-3}$). 


The middle and right panels of Figure~\ref{fig:feh_dist_ns_gc} show a comparison of the cumulative distributions of total energy and $z$-component of the angular momentum of the high-[N/O] stars and GCs with metallicities $\rm [Fe/H]<-1$. It clearly shows
that distributions of the energy and angular momentum of the in-situ stars and GCs agree, while in-situ stars are inconsistent with the distributions of ex-situ GCs both in metallicity and energy. 

We also find that distributions of metallicity, energy, and angular momentum of the ex-situ high-[N/O] stars and ex-situ GCs are consistent. The uncertainties in this comparison are larger, however, as there are only 25 ex-situ high-[N/O] stars in our sample. Still,  
distribution of ex-situ stars and in-situ stars is strongly inconsistent in energy (probability to be drawn from the same distribution is 0.0), which is not surprising because boundary between in-situ and ex-situ objects indicated by $\rm [Al/Fe]$ corresponds to a rather sharp boundary in energy.

Our results appear to be inconsistent with results of a recent study by \citet{Kim.etal.2023} who also find that N-rich stars have metallicity distribution similar to that of GCs at $\rm [Fe/H]<-1$, but who argued that distributions of orbital parameters of these stars and clusters
are not the same. We note, however, that in their Figure~4 only distributions of the pericenters ($r_{\rm min}$) are somewhat different. 
Furthermore, if similarity exists only for the low-metallicity ($\rm [Fe/H]<-1$) regime and the goal is to test whether similarity exists also in the orbital properties of stars and clusters, it would be more conclusive to do a comparison of orbital
properties in that regime, as we do here, rather than for the entire sample.

Overall, our results presented in Figure~\ref{fig:feh_dist_ns_gc} shows  that distributions of metallicity, energy, and momentum of the in-situ GCs and in-situ high-[N/O] stars with $\rm [Fe/H]<-1$ are similar, {\it which strongly indicates that low-metallicity in-situ globular clusters are the primary formation sites of these stars.} 

If we assume that this is the case, the observed fraction of high-[N/O] stars in the Aurora population allows us to estimate the fraction of stars born in clusters during early, pre-disk stages of  MW evolution. We present such an estimate in the next subsection.

% \begin{figure*}
%   \centering
%   \includegraphics[width=\textwidth]{img/gcns_feh_elz_dist.pdf}
%   \caption[]{Cumulative distributions of metallicity ([Fe/H]) of the in-situ $N$-rich stars (gray solid line) and in-situ globular clusters (blue dashed line). The gray and blue shaded regions show corresponding 95\% range estimated using bootstrap resampling. The two samples were selected to have [Fe/H]$<-0.5$ and using our adopted criteria for in-situ populations. The figure shows that distribution of the in-situ GCs and $N$-rich clusters are remarkably similar, especially at [Fe/H]$<-1$.}
%    \label{fig:feh_dist_ns_gc}
% \end{figure*}

\subsection{The fraction of the Aurora stars born in clusters}
\label{sec:fcl}

The correlation between the fraction of high-[N/O] stars and the initial GC mass shown in the middle panel of Figure~\ref{fig:gc_mass} and estimates of the $f_{\rm N/O}$ at different metallicities (Figure~\ref{fig:frac_feh2}) can be used to compute  $f_{\rm N/O,cl}$ -- the fraction of stars born in clusters that have high N/O for a population of newly born GCs in high-$z$ MW progenitor. This, in turn, can be used to estimate the fraction of stellar mass born in clusters in the Aurora population.

To this end, we adopt a model for the mass function of clusters at birth that follows the Schechter functional form \citep[][]{Schechter1976}: $dN/dM \propto M^{-\beta} e^{-M/M_c}$.  
We choose the cut-off mass of $M_c=5\times10^6 M_{\odot}$ motivated by the analysis of \citet{Choksi2019} and observational estimates \citep[see Fig. 17 in][]{Adamo.etal.2020}, as well as  $\beta=2$, the average slope of the mass function of young compact star clusters measured in observed galaxies at $z=0$ \citep[see][for a review]{Krumholz2019}, and assume that clusters are born with masses in the range $10^3-5\times10^7\ M_{\odot}$ \citep[see, e.g.,][for an observational justification of the upper value]{Norris.etal.2019}. The resulting estimate using such a model gives the fraction of high-[N/O] stars born in clusters  $f_{\rm N/O,cl}\approx14\%$. Increasing characteristic mass to $M_c=10^7M_{\odot}$ and the minimal cluster mass to $10^4M_{\odot}$ while varying the $f_{\rm N/O}-M_{\rm ini}$ within the grey band shown in the middle panel of Figure~\ref{fig:gc_mass} gives a range of typical high-[N/O] fractions $6\%<f_{\rm N/O,cl}<32\%$. 

Given this estimated range and the observed fraction of high-[N/O] stars in the Aurora population, we can estimate the fraction of the Aurora stars born in clusters as $f_{\rm cl}\approx f_{\rm N/O}/f_{\rm N/O,cl}$. The MW in-situ stars with metallicities of $\rm [Fe/H]\approx -1.5$ exhibit $f_{\rm N/O}\approx 4\%$.
Thus, these fractions indicate that the fraction of the Aurora stars born in clusters ranges from $\approx 15\%$ to $\approx 70\%$. 
The range corresponding to 50\% of all stars born in clusters is shown as a horizontal gray band in the middle panel of Figure~\ref{fig:frac_feh2}.   
Aurora stars of higher metallicity $\rm [Fe/H]\approx -1$ have $f_{\rm N/O}\approx 1\%$ and fractions of stars born in clusters is correspondingly four times smaller: $\approx 4-17\%$. These estimates can be compared to the GS/E progenitor, where a considerably smaller fraction of $2-10\%$ is estimated to be born in clusters independent of metallicity. 

Note that our estimate of $f_{\rm N/O,cl}$ implicitly assumes that all of the stellar mass initially in GCs is now in the field. 
This is supported by modelling results of \citet[][see their Fig. 3]{Rodriguez2023} and \citet[][see their Fig. 13]{Gieles2023} which indicate that the majority of the clusters contributing large numbers of high-[N/O] stars disrupt by $z=0$. Furthermore, according to the initial mass estimates for the surviving in-situ clusters by \citet{Baumgardt2003}, that population lost $\approx 84\%$ of its initial combined mass. Thus, within the uncertainties of our estimate, the fraction of stellar mass born in GCs that is still in surviving clusters is negligible.  

%-----------------------------------------------------
\subsection{Estimate of mass of the Aurora stellar population}
\label{sec:maurora}
%-----------------------------------------------------

The estimate of mass fraction of the in-situ stars born in clusters presented above can be used 
to estimate the stellar mass of the Aurora population -- the in-situ stars with metallicities $\rm [Fe/H]\lesssim -1$. This is useful because MW globular cluster propulation is more or less complete, while Aurora stars are observed over a limited volume in a sample
with a fairly uncertain selection function. 

Indeed, this mass can be estimated as
\begin{equation}
M_{\rm Aur}=\frac{1}{1-f_{\rm disrupt}}\,\sum_{i=1}^{N_{\rm GC,ins}}\, \frac{M_{i,\rm ini}}{f_{i,\rm cl}},
\label{eq:maurora}
\end{equation}
where $f_{\rm disrupt}$ is the fraction of the initial clusters that were disrupted, $N_{\rm GC,ins}$ is the number of observed in-situ clusters with metallicity $\rm [Fe/H]< -1$, $f_{i,\rm cl}({\rm [Fe/H]})$ is the fraction of star formation that occurred in bound star clusters at metallicity $\rm [Fe/H]_i$, and $M_{\rm i,ini}$ is the initial mass of the observed cluster $i$, respectively. We use the initial GC mass estimates from \citet{Baumgardt2003} and approximation to $f_{\rm cl}({\rm [Fe/H]})=f_{\rm N/O}({\rm [Fe/H]})/0.14$, where $f_{\rm N/O}({\rm [Fe/H]})$ is approximation to the trend shown in Figure~\ref{fig:frac_feh}:
\begin{equation}
f_{\rm N/O} = \frac{a}{(1+x^b)^{g/b}},    
\label{eq:fno_approx}
\end{equation}
where $x=10^{\rm [Fe/H]+1.2}$, $a=3\times 10^{-2}$, $b=6$, $g=3.25$.

Using eqs~\ref{eq:maurora} and \ref{eq:fno_approx} and our classification of the in-situ clusters in the 
\cite{Baumgardt_Vasiliev2021} catalog, we estimate stellar mass of the Aurora population $M_{\rm Aur}\approx 4.6\times 10^8\, M_\odot$. The uncertainty of this estimate is due to uncertainty 
in $f_{\rm cl}({\rm [Fe/H]})$, the estimate of the fraction of high-[N/O] stars in clusters, the uncertainty in the estimate of initial cluster masses and uncertainty in the fraction of initial globular cluster population that was disrupted before $z=0$. 
A rough estimate of the first uncertainty can be obtained by changing the value of constant $a$ in equation~\ref{eq:fno_approx} from $2\times 10^{-2}$ to $4.5\times 10^{-2}$ to roughly encompass the uncertainties of $f_{\rm N/O}$ measurements in Figure~\ref{fig:frac_feh}. This gives estimates of the Aurora population mass ranging from $3\times 10^8\, M_\odot$ to $6.9\times 10^8\, M_\odot$. Thus, the mass of this population is approximately

\begin{equation}
  M_{\rm Aur} = 5\pm 2 \times 10^8 (1-f_{\rm disrupt})^{-1}\, M_\odot,
  \label{eq:maur_est}
\end{equation}
although the uncertainty in this estimate only includes uncertainty in $f_{\rm N/O}$ and not the other sources listed above. The uncertainty in the initial cluster mass estimate is a systematic uncertainty of this estimate. The unknown fraction of disrupted clusters is represented by
the factor $1-f_{\rm disrupt}$ accounts for the fact that we use surviving clusters to make the estimate, while $f_{\rm disrupt}$ fraction of the initial mass in cluster was returned to the field due to cluster disruption. GC evolution discussed in the previous section indicate that the fraction of disrupted clusters is $f_{\rm disrupt}\approx 2/3$. The actual mass of the Aurora population is thus larger by the corresponding factor than the above estimate. 

\begin{figure}
  \centering
  \includegraphics[width=0.49\textwidth]{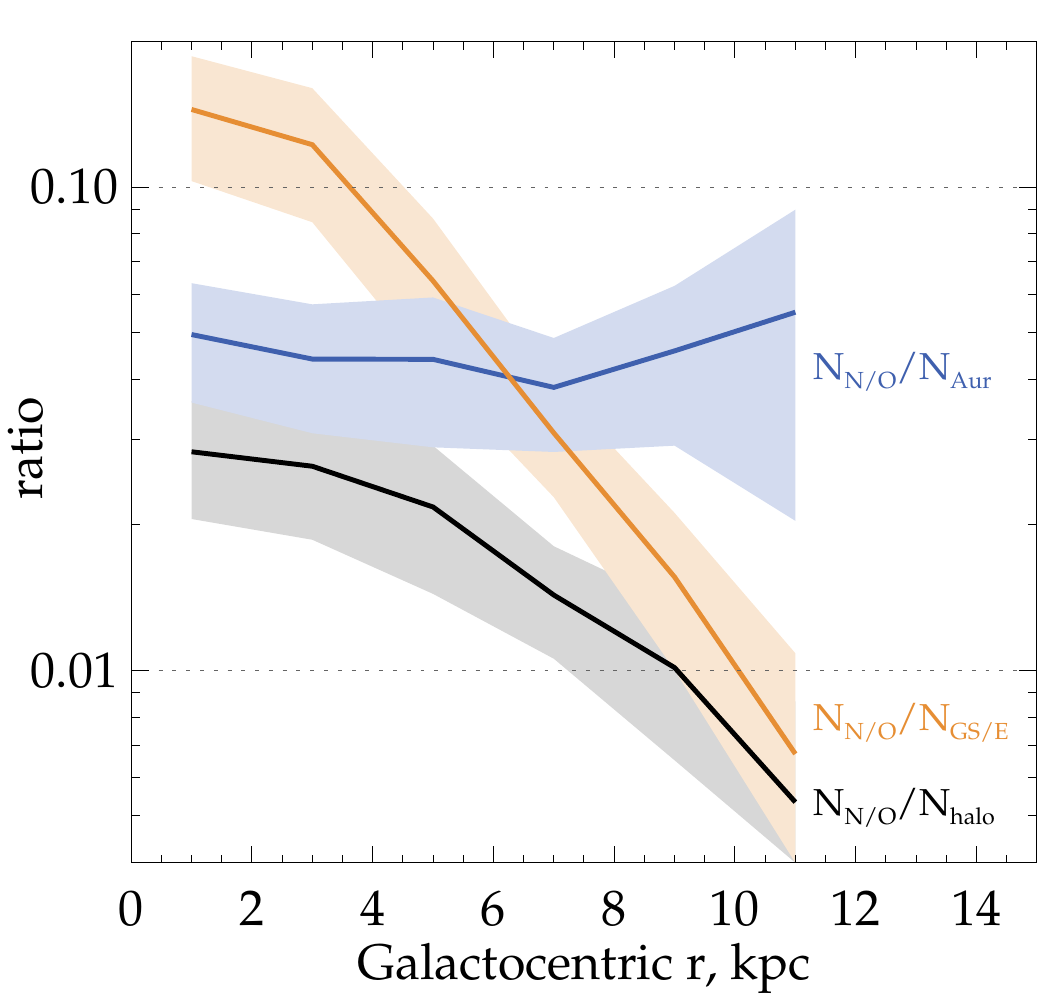}
  \caption[]{Ratios of the number of high-[N/O] stars to the number of stars in different Galactic components as a function of Galactocentric radius. Only stars with $-1.45<$[Fe/H]$<-1$ are considered. Grey shows the ratio (and the associated uncertainty) relative to all halo stars as a function of $r$. The halo fraction of high-[N/O] stars increases by a factor of $\approx6$ from the Solar radius to the Galactic centre, in agreement with results of \citet{Horta2021}. The orange band shows the ratio of high-[N/O] stars to the accreted halo population (mainly GS/E) and exhibits a much steeper, a factor of $\approx 40$, increase. This is because the GS/E debris does not populate the inner regions of the MW and thus has a shallower density profile compared to the halo overall and Aurora in particular. The blue band is the ratio relative to the Aurora stars. This ratio is flat at $\approx4\%$, indicating that the high-[N/O] stars in this metallicity range belong to the Aurora population.}
   \label{fig:ratio_rad}
\end{figure}

Notwithstanding the uncertainties, this estimate shows that the mass of the in-situ low-metallicity population of stars that were born before MW disk started to form and that have now a spheroidal distribution is comparable to the mass of the MW stellar halo of $\approx 1.4\pm 0.4\times 10^9\, M_\odot$ \citep[e.g.,][]{Deason.etal.2019}. Given that this in-situ population is more concentrated towards the centre of the Galaxy than the overall halo (\citealt{Rix2022}; see comparison of the distance distributions the Aurora stars and halo stars in Figure~\ref{fig:ratio_rad}), it is expected to dominate the spheroidal stellar component at $r\lesssim 10$ kpc. 

By definition the Aurora population is the population of in-situ stars formed before the spin-up of the Milky Way disk. Its mass estimate above is thus also the estimate of the Milky Way progenitor's stellar mass at the time when disk started to form and indicates that this occurred when Milky Way's stellar mass was between the mass of the Small and Large Magellanic Clouds. 

\subsection{Radial distribution of high-[N/O] stars}

\citet{Horta2021} show that the fraction of the fraction of N-rich stars among halo stars increases by a factor of about six  from $r\approx 10$ kpc to $r=2$ kpc. They put forward a hypothesis that such an enhancement of GC-born stars may be linked with an increase in accretion and disruption of dwarf galaxies in the high-$z$ Milky Way. Note, however, that a much flatter trend out to $\sim40$ kpc is reported \citep[see][]{Koch2019,Horta2021}.

Figure~\ref{fig:ratio_rad} explores the radial dependence of the fraction of high-[N/O] stars in the Galactic halo (black line, grey band). We also compute the ratio of the number of high-[N/O] stars to the number of stars in the GS/E ($N_{\rm N/O}/N_{\rm GS/E}$, orange) and Aurora ($N_{\rm N/O}/N_{\rm Aur}$, blue) components. Here we resort to the [Al/Fe]-based in-situ/accreted classification as the energy-based selection will induce a bias in the radial distribution. Therefore, the radial profiles shown in Figure~\ref{fig:ratio_rad} are limited to stars with $-1.4<$[Fe/H]$<-1.1$, where [Al/Fe] selection is most reliable (see discussion above). As was shown in the right panel of Figure~\ref{fig:frac_feh}, the difference for samples selected by energy and [Al/Fe] is much smaller than the trends we are considering. 

While the details of our selection of high-[N/O] stars are different to that in \citet{Horta2021}, the halo fraction in these stars shows a very similar trend with Galacto-centric radius to the one estimated in that study. As the black curve indicates, there are $\sim6$ times more high-[N/O] GC-like halo stars in the centre of the Galaxy compared to their relative counts around the Sun. A much more dramatic radial change in the fraction of high-[N/O] stars can be seen when their numbers are compared to the counts of the accreted debris (composed mostly of the GS/E stars). Relative to the GS/E stars, the high-[N/O] population increases by a factor of $\sim 25$ from 10 kpc to 1 kpc. 

This is in agreement with recent studies of the inner stellar halo which indicate that the fraction of the GS/E stars decreases within $\sim3$ kpc from the Galactic centre \citep[see][]{Iorio2021,wrinkles}. The Galacto-centric radial distribution of the tidal debris from an accteted satellite depends on the mass of the dwarf and the merger time \citep[see][]{Deason2013,Horta_FIRE}. Note that the GS/E progenitor galaxy was likely of sufficient mass to have experienced strong dynamical friction and profound loss of the orbital energy and angular momentum during its interaction with the Milky Way \citep[see discussion in][]{radialize}. Such rapid orbital radialization is often accompanied by "explosive" mass loss resulting in a complete disruption of the satellite before it manages to reach the centre of the Galaxy. The drop in relative density of the GS/E debris inward of the Solar radius can also be gleaned from Figure~\ref{fig:elz_aurora} where the number of orange points quickly decreases below the Sun's energy level.

Large changes in the relative fraction of high-[N/O] stars when compared to the stellar halo overall or to its accreted component signal one thing: high-[N/O] GC-like population is not a typical member of either of these. Instead, as shown by the blue curve in Figure~\ref{fig:ratio_rad}, the radial distribution of the high-[N/O] stars is very similar to that of the Aurora stars and they are thus likely to be a component of the Aurora population. This is in complete agreement with the above comparisons between the trends of the numbers of high-[N/O] and Aurora stars as a function of energy (Figure~\ref{fig:elz_aurora}) and metallicity (Figure~\ref{fig:frac_feh}). 

Turning the argument around, the above discussion cements the view of the Aurora population as a centrally concentrated stellar halo component, as hypothesised by BK22 based on a dataset limited to the Solar vicinity. Such a radial density (strongly peaked at $r=0$) and an extent (limited to approximately Solar radius) is in agreement with the results of numerical simulations of the Milky Way formation as illustrated in Figure 13 of BK22 and observed in the all-sky view of the metal-poor component of the Galaxy \citep[see][]{Rix2022}. These recent studies \citep[see also][]{Conroy2022,Myeong2022} together with the trends discussed here indicate that at low metallicities, the central stellar halo is dominated by the in-situ formed Aurora stars rather than by stars brought in by other accreted galaxies.

\begin{figure}
  \centering
  \includegraphics[width=0.495\textwidth]{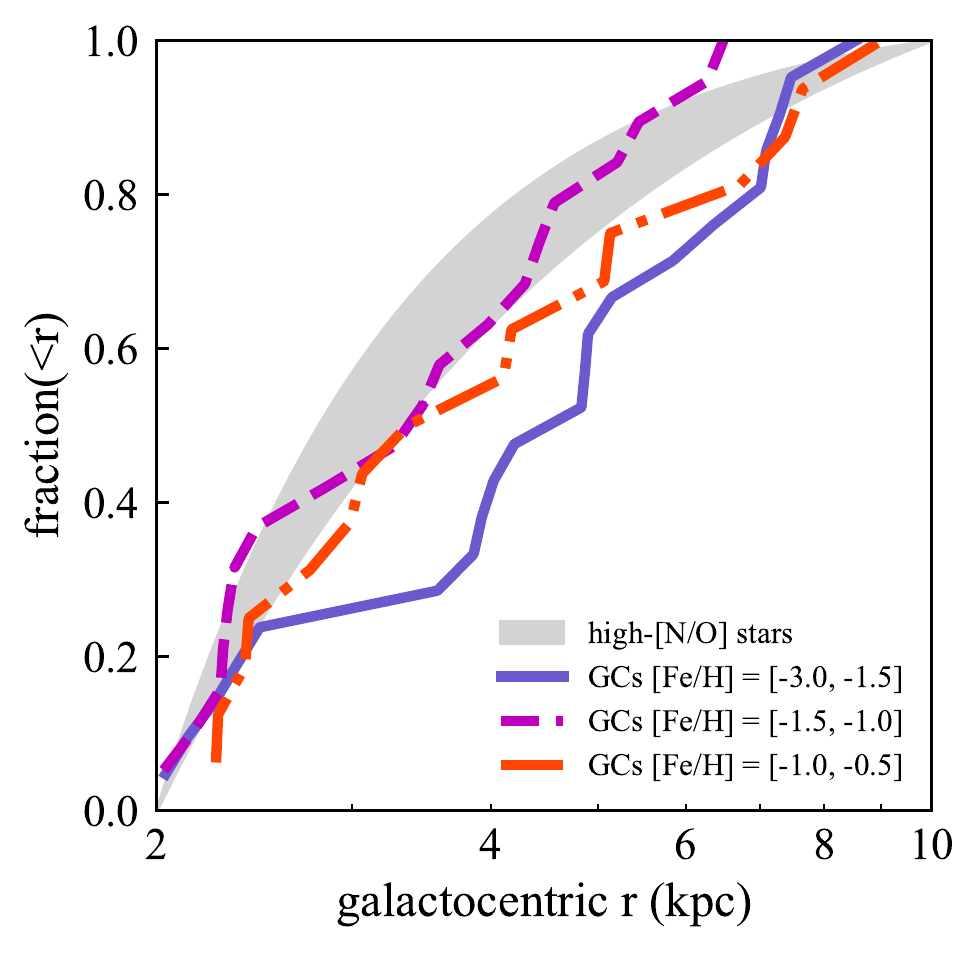}
  \caption[]{Cumulative distribution of the galactocentric distance for the MW in-situ GCs in three metallicity intervals (thick lines) and radial distribution of the $N$-rich stars from \citet{Horta2021} with $\rm [Fe/H]<-1$ shown by the grey band. The band represents the uncertainty in the slope of the derived profile of $N$-rich stars. Given that the distribution of the high-[N/O] stars was derived for the range $d_{\rm GC}=[2,10]$ kpc only GCs in the same distance range are included in the cumulative distribution. The plot shows that distance distributions of the high-[N/O] stars and in-situ GCs with $\rm [Fe/H]>-1.5$ are consistent indicating that these stars may have originated from the in-situ clusters in this metallicity range, which is also where metallicities of most high-[N/O] stars in the \citet{Horta2021} sample lie.  } 
   \label{fig:dgc_nrich_dist}
\end{figure}

Given the indications that high-[N/O] stars originated in clusters that we discussed above, it is also instructive to compare the radial distribution of these stars with the radial distribution of in-situ globular clusters. Such comparison is shown in Figure~\ref{fig:dgc_nrich_dist} where the cumulative
distribution of the galactocentric distance of high-[N/O] stars estimated by \citet{Horta2021} is compared to cumulative distance distributions of the in-situ globular clusters in our classification in [Fe/H] ranges $[-3,-1.5]$, $[-1.5,-1]$ and $[-1,-0.5]$. Cumulative distributions for both clusters and high-[N/O] stars were constructed using objects with distances in the range $r\in[2-10]$ kpc because this was the distance range of the \citet{Horta2021} sample.

% %

The figure shows that the distance distribution of GCs with $\rm [Fe/H]>-1.5$ is quite similar to that of the high-[N/O] stars. In particular, the distributions are closest for clusters with metallicities $\rm [Fe/H]\in[-1.5,-1]$, which corresponds to the range of metallicities of most of the high-[N/O] stars. Recently, \citet{Gieles2023} have also shown that the distance distribution of stars from disrupted massive GCs in their model matches the distribution of N-rich stars (see their Section 6.2 and Figure 12) and argued that this is an indication that these stars originated in clusters. In principle, the distance distribution of the surviving clusters and disrupted clusters can be different, but our results indicate the difference is not large. 
The largest difference between distance distributions in Figure~\ref{fig:dgc_nrich_dist} is for in-situ GCs with $\rm [Fe/H]<-1.5$, although the difference is not very significant as the number of clusters in this metallicity range is fairly small. 

\begin{figure}
  \centering
  \includegraphics[width=0.49\textwidth]{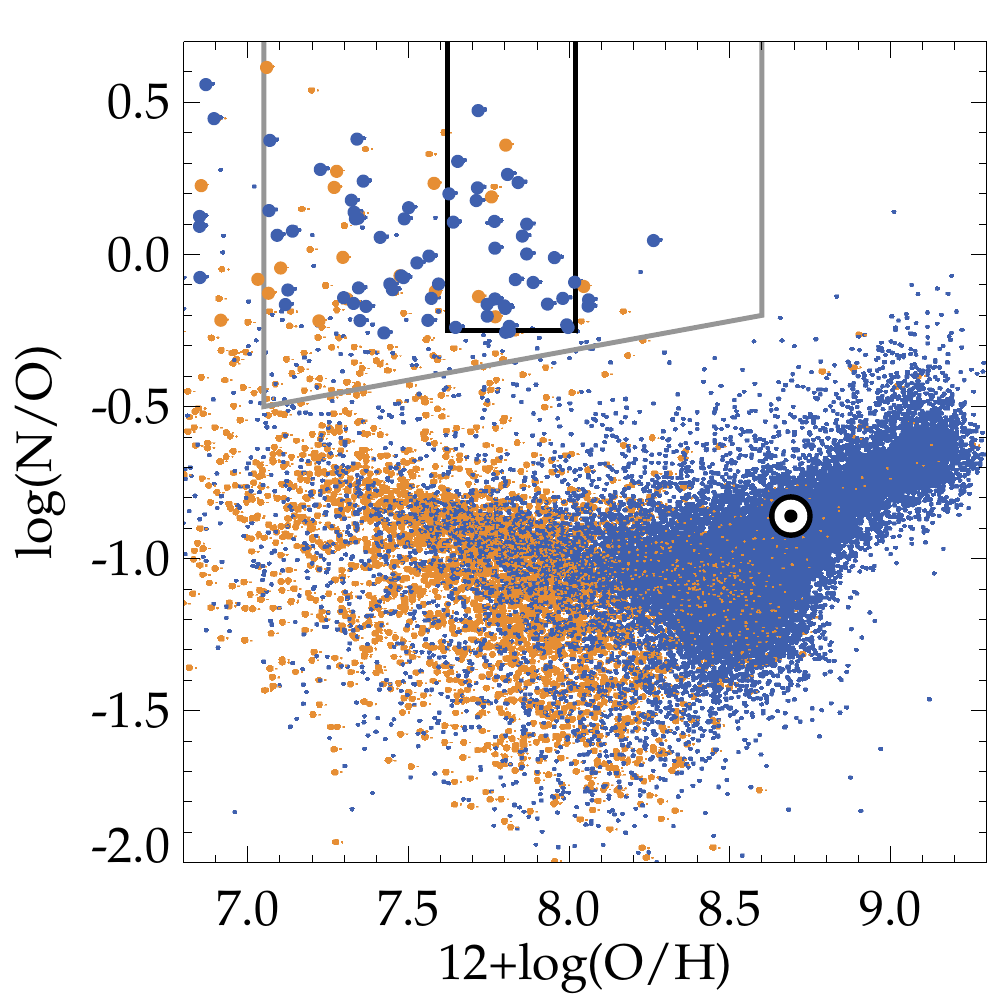}
  \caption[]{Nitrogen-to-oxygen $\log({\rm O/H})=[{\rm N/O}]+\log({\rm N/O})_{\odot}$ abundance ratio as a function of oxygen abundance $12+\log({\rm O/H})=[{\rm O/H}]+\log({\rm O/H})_{\odot}$ for giants with low $V_\phi$ classified as accreted (orange) and those born in-situ (blue, including Aurora). Stars are classified into accreted and in-situ based on their energy $E,L_z$ values. Conservative (thin black lines) and fiducial (thick black lines) abundance measurements for GN-z11 from \citet{Cameron2023} are also shown. Solar $\log({\rm N/O})_{\odot}=-0.86$ and $12+\log({\rm O/H})=8.71$ is marked with $\odot$. Giants with high N/O ratio fall squarely within the range of measurements for GN-z11. Note that in accordance with Fig~\ref{fig:elz_aurora}, at least 75\% of the high-[N/O] stars belong to Aurora.} 
   \label{fig:no_oh}
\end{figure}

The similarity of metallicity, energy, and angular momentum distributions of high-[N/O] stars in clusters shown in Figure~\ref{fig:feh_dist_ns_gc} and the similarity of distance distributions of these stars with the Aurora stars and in-situ GCs shown in Figures~\ref{fig:ratio_rad} and \ref{fig:dgc_nrich_dist} strongly indicate that  {\it the majority of high-[N/O] stars with $\rm [Fe/H]<-1$ originated in globular clusters formed in-situ in the low-metallicity Aurora population.}  

\subsection{High-redhsift Milky Way vs GN-z11}
\label{sec:gnz11}

Figure~\ref{fig:no_oh} presents the [N/O] abundance ratio as a function of the oxygen-based metallicity [O/H]. This view is similar to that shown in panel c of Figure~\ref{fig:select}, albeit here we show the absolute abundance ratios, i.e. not referenced to the solar values. Instead, the solar values are marked with the $\odot$ symbol. Below $12+\log({\rm O/H})\approx8.2$ both accreted (orange) and in-situ (blue) populations are present. Above $12+\log({\rm O/H})>8.2$ the stars are overwhelmingly of in-situ origin. Up to the Solar value of $12+\log({\rm O/H)=8.69}$, the N/O is almost flat with a small bend. Around the Solar metallicity, the N/O abundances start to rise noticeably. Conservative (fiducial) measurements reported by \citet{Cameron2023} for the GN-z11 galaxy are shown with grey (black) lines. 

The Figure shows a large number of APOGEE field giants with N/O and O/H abundance ratios similar to those reported for GN-z11. There is a total of 100 high-[N/O] low-metallicity ($\rm [Fe/H]<-0.7$) stars marked with larger filled circles. Of these 73 are classified unambiguously as in-situ and 27 as accreted. 
Given our conclusion that the majority of the in-situ high-[N/O] stars originated in massive bound clusters and that the clusters, in turn, constituted a significant fraction of star formation at early epochs, the implication of these
results is that a significant fraction of star formation in GN-z11 also occurs in bound compact star clusters. 

\begin{figure*}
  \centering
  \includegraphics[width=0.49\textwidth]{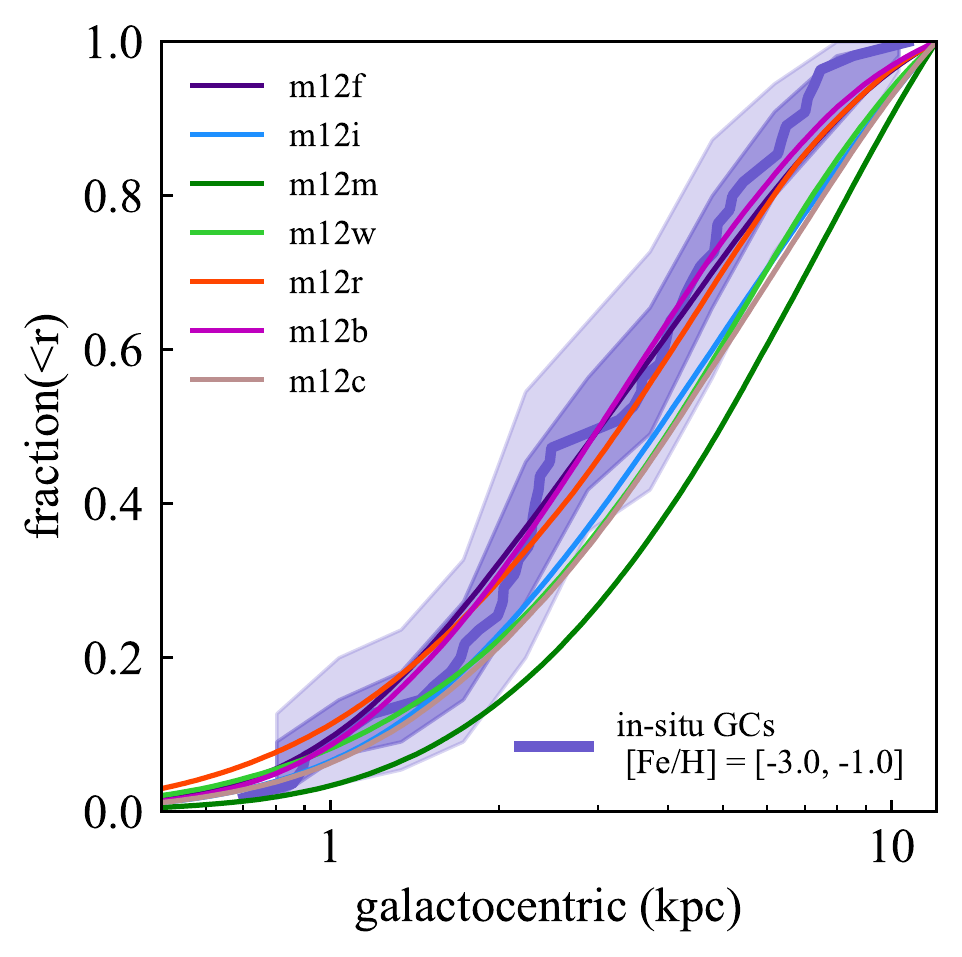}
  \includegraphics[width=0.501\textwidth]{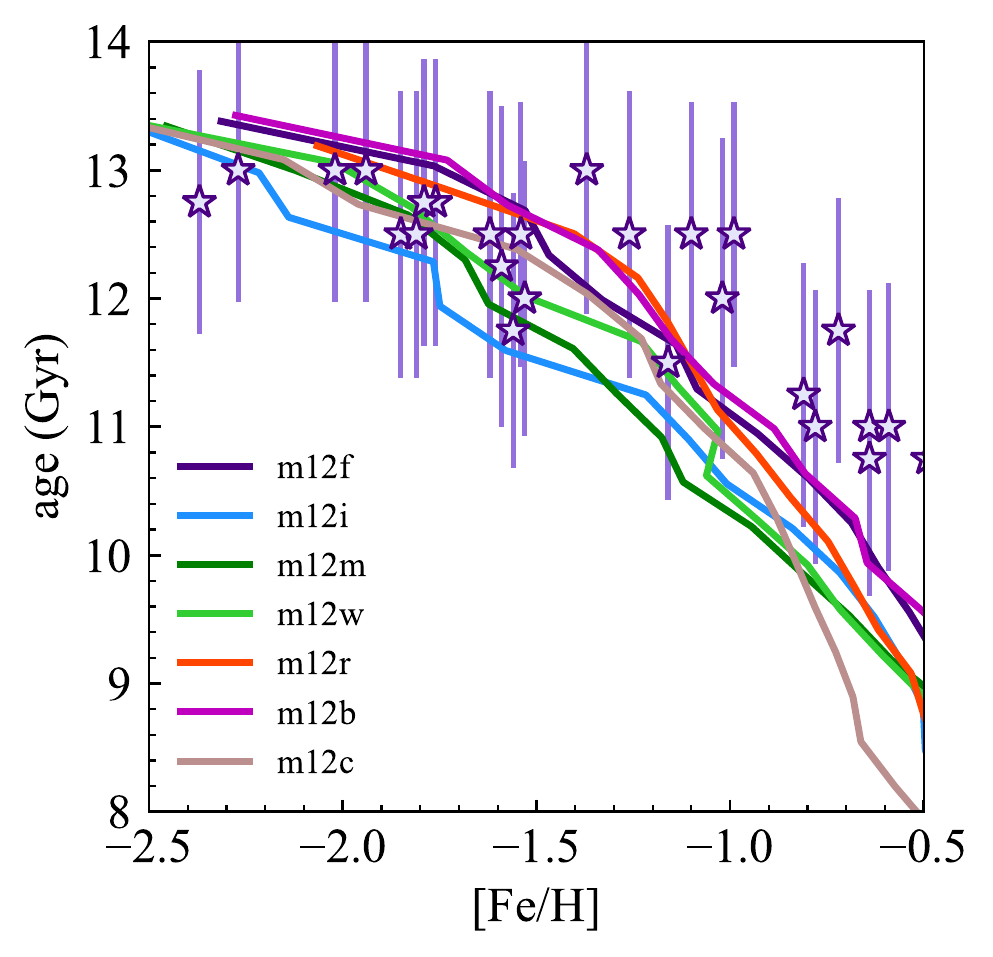}
  \caption[]{{\bf Left:}\/ Cumulative distribution of the galactocentric distance for the MW in-situ GCs (thick lines with shaded bands) in the metallicity interval $\rm [Fe/H]\in [-3,-1]$ 
  compared to the distribution of galactocentric distance of the in-situ born stars in the FIRE simulations of the MW-sized galaxies in the same metallicity interval. The figure shows that the distribution of observed in-situ clusters and in-situ stars in a significant fraction of simulated galaxies agree indicating that MW GCs could form as part of the regular in-situ star formation in the early Milky Way. {\bf Right}: age-metallicity relation for the in-situ stars in simulations and in-situ MW GCs.} 
   \label{fig:dgc_insitu_dist}
\end{figure*}

\section{Discussion}
\label{sec:discussion}

Results presented above strongly indicate that 1) the majority of low-metallicity high-[N/O] stars originated in compact bound star clusters and 2) up to $f_{\rm cl}\approx 50\%$ of star formation at early epochs before MW formed its disk (${\rm [Fe/H]}\lesssim -1.5$) occurred in bound massive star clusters. This fraction decreases rapidly with increasing metallicity to $10-20\%$ at ${\rm [Fe/H]}\approx -1$ and to $\lesssim 1\%$ at ${\rm [Fe/H]}\gtrsim -0.7$.

A number of recent studies have estimated the fraction of field N-enhanced stars in the Milky Way and showed it to be small, of the order of $\sim2\%$ \citep[e.g.,][]{Martell2016, Schiavon2017, Koch2019, Horta2021}, consistent with the measurements presented here. Some of these studies 
have also used the estimated fraction to derive the fraction of low-metallicity stars that formed in bound clusters, obtaining values $f_{\rm cl}\approx 5-20\%$  \citep[e.g.,][]{Martell2011,Koch2019,Horta2021}. 

As we discussed in Section~\ref{sec:intro}, these estimates involve the total stellar population without accounting for the contribution of in-situ and ex-situ stars. Thus, the estimated fraction is averaged over stars formed in the accreted dwarf galaxies and the main MW progenitor, which likely had very different environments and evolution. Interpretation of such an average is not straightforward. Furthermore, the estimated $f_{\rm cl}$ uses an uncertain factor to convert the estimated fraction of N-enhanced stars into the fraction of the first generation stars in clusters. Finally, the threshold used to select N-rich stars is somewhat arbitrary and is not calibrated on the real GCs, without correction to their initial mass, and without accounting for the GC mass function. In this study, we estimate $f_{\rm cl}$ rectifying all of these issues.  

Estimates of the contribution of the UV luminosity from the young progenitors of modern globular clusters to the total UV luminosity function of galaxies \citep{Ricotti.2002,Boylan.Kolchin.2018} provide an 
independent line of evidence that compact massive clusters were a significant fraction of star formation at $z\gtrsim 4$. Indeed,
direct estimates using observed globular cluster in the ultra-faint galaxy Eridanus II \citep{Weisz.etal.2023} show that the cluster contributed up to $\approx 10\%$ of the galaxy stellar mass at birth, while \citet{Zick.etal.2018} show that $\approx 20\%$ of star formation in the Fornax dwarf occurred in bound {\it surviving} clusters. Similarly, 
we estimate that for the GS/E the fraction of stellar mass formed in GCs is $f_{\rm cl}\approx 0.06\,f_{\rm GC,GS/E}\,M^{-1}_{\star,8.7}$, where $M^{-1}_{\star,8.7}$ is GS/E's stellar mass at the time of the merger in $5\times 10^8\, M_\odot$ and $f_{\rm GC,GS/E}$ is the fraction of MW's ex-situ clusters contributed by GS/E. However, $\approx 55\%$ of ex-situ GCs mass is in clusters with $\rm [Fe/H]\leq -1.5$, while only $\approx 15\%$ of GS/E stars are at these metallicities (see lower panel of Fig. 4 in BK22). If we assume that all ex-situ GCs were formed in GS/E, this corresponds to $f_{\rm cl}\approx 25\%$ in the GS/E at $\rm [Fe/H]\leq -1.5$. 

We also see a direct record of GC formation in the linear relation between mass in globular cluster population in galaxies and mass of the parent halo, $M_{\rm GCS}=\eta_{\rm GC}M_{\rm h}$. Such relation is expected in the models of GC formation \citep{Kravtsov2005,Bekki.etal.2008, Choksi2018,Choksi2019,El_Badry_etal.2019} and is observed to hold over more than five orders of magnitude in galaxy stellar mass with $\eta_{\rm GC}\approx 3\times 10^{-5}$ \citep[see][]{Spitler.Forbes.2009,Hudson.etal.2014,Harris.etal.2017,Forbes.etal.2018,Dornan.Harris.2023}. Given that stellar mass--halo mass relation is nonlinear and in the stellar mass range of the MW progenitors is close to $M_\star\propto M_{\rm h}^\alpha$ with $\alpha\approx 1.5-2$ for $M_{\rm}\approx 10^{11}-10^{12}\, M_\odot$ and $\alpha\approx 2-2.5$ for smaller masses \citep[e.g.,][]{Kravtsov.etal.2018,Read.Erkal.2019,Nadler.etal.2020}, the fraction of stellar mass in bound star clusters increases with decreasing stellar mass as $f_{\rm cl}\propto M^{1/\alpha-1}_\star$, at least for galaxies of $M_\star\gtrsim 10^7\, M_\odot$ that contain at least a single GC \citep[e.g.,][]{Chen_Gnedin2023}. We thus can expect the mass fraction of stars formed in bound clusters to increase in with decreasing stellar mass and metallicity for $M_\star\lesssim 10^{10}\, M_\odot$ and variations of the fraction of star formation in bound clusters are therefore quite ubiquitous. 
 
High-$z$ galaxies with a significant fraction of rest-frame UV light concentrated in compact (sizes of tens of parsecs) clumps
are also directly observed \citep[e.g.,][]{Livermore.etal.2015,Johnson.etal.2017,Vanzella.etal.2023}, although such observations are few because they require a lucky lensing configuration.
At $z\approx 0$, however, the fraction of UV light in compact bound clusters is found to be a function of star formation surface density,  $\Sigma_{\rm SFR}$, \citep[e.g.,][see also \citealt{Adamo.etal.2020rev}  for a review]{Adamo.etal.2020} reaching $\sim 50-100\%$ in galaxies with highest $\Sigma_{\rm SFR}$. The overall observed trend is qualitatively consistent with expectations of cluster formation models \citep{Kruijssen2012}. Furthermore,  during early epochs galaxy progenitors are 1) in the fast mass accretion regime, 2) have small sizes, and 3) are expected to have high gas fractions. We can thus expect 
that surface density of gas and star formation, and correspondingly the fraction of star formation in bound clusters, to be comparably high during these epochs. 
Observations of both local and high-$z$ galaxies are therefore consistent with the conclusion that a significant fraction of star formation in the early MW occurred in compact, bound star clusters. 

The conclusion that a significant fraction of stars formed in compact bound clusters also has a straightforward implication for galaxy formation simulations. If modern simulations do a good job modelling stellar populations of observed galaxies \citep[see][for reviews of recent dramatic progress in such simulations]{Naab.Ostriker.2017,Vogelsberger.etal.2020} and if GCs were a significant fraction of early star formation, we can expect that spatial distribution of the low-metallicity in-situ stars born in simulations should be similar to that of observed GCs. 

Indeed, the left panel of Figure~\ref{fig:dgc_insitu_dist} shows a remarkable similarity between the galactocentric distance distribution of the low-metallicity in-situ stars in the FIRE-2 simulations \citep{Hopkins.etal.2018,Wetzel.etal.2023} of the MW-sized progenitors and 
the corresponding distribution of the MW GCs. Although the model distribution varies from object to object, the predictions are quite close to the observed GC distribution and the variations are much smaller than, say, the difference in the distance distribution of the MW in-situ and accreted GCs. In fact, six out of seven of the hosts have predicted distance distributions within the bootstrap $2\sigma$ uncertainty of the MW GC distribution (the outer shaded band).  

Moreover, as can be seen in the right panel of Figure~\ref{fig:dgc_insitu_dist}, the FIRE-2 galaxies that match the distance distribution of the MW GCs ({\tt m12b, m12f, m12r}) also have the age-[Fe/H] distribution that is closest to that of the MW GCs. These hosts have the earliest formation (oldest ages of stars at a given metallicity), which indicates that MW star formation history corresponds to the earliest-forming tail of the MW-sized galaxies. This is 
consistent with independent indications of the early formation of the MW based on the occurrence and timing of the GS/E merger discussed by \citet{Fattahi.etal.2019} and \citet{Dillamore.etal.2022}.

Interestingly, if we adopt $f_{\rm N/O}$ as a function of [Fe/H] estimated in this study (see Fig.~\ref{fig:frac_feh2} and eq.~\ref{eq:fno_approx}) as a proxy for the functional form of
the $f_{\rm cl}(\rm [Fe/H])$ dependence, and convolve the metallicity distribution of in-situ stars in the FIRE-2 simulation with that function, the resulting distribution of metallicities has a broad peak at $\rm [Fe/H]\sim -1\div -1.5$ resembling broadly metallicity distribution of the low-metallicity in-situ MW GCs. This consistency check indicates that the assumption $f_{\rm cl}\propto f_{\rm N/O}$ is broadly consistent with observed properties of the MW in-situ GCs.

The plausible models for producing enhanced abundances of nitrogen involve massive stars \citep[see, e.g.,][]{Bastian_Lardo2018} -- massive rotating stars \citep[$M\gtrsim 15\, M_\odot$, e.g.,][]{Maeder.Meynet.2006,Crowther.2007} or very massive stars of $\sim 10^3-10^4\, M_\odot$ \citep{Denissenkov.Hartwick.2014}.  If stellar collisions or some other physical processes that facilitate the formation of massive stars at low metallicities in clusters can effectively create a substantial number of massive and very massive stars, this can potentially make the initial mass function (IMF) more top-heavy \citep[see, e.g.,][]{Dib.etal.2007,Chon.etal.2021} and significantly boost the UV luminosity per unit of stellar mass formed. 

Even without a significant change to the IMF, a significant fraction of star formation in bound clusters will result in many short-duration bursts of star formation when each cluster forms. 
Clusters older than $\sim 3-5$ Myrs are usually observed to be unobscured with an ionized environment, which means that they form on the time scales $\lesssim 3$ Myrs. For a galaxy like GN-z11 with a stellar mass $M_\star\approx 5\times 10^8\, M_\odot$ at $z\approx 10.6$ corresponding to the cosmic time of $\approx 430$ Myr, $f_{\rm cl}=0.5$ implies that $\sim 100-200$ bound clusters of mass $\sim 10^6-10^7\, M_\odot$ form during $\sim 300-400$ Myr. On Myr time scales the UV luminosity of the parent galaxy will thus spike shortly after cluster formation and will rapidly decrease before the next cluster forms. Thus, as the stellar mass grows the $M_\star/L_{\rm UV}$ ratio will fluctuate strongly due to individual cluster formation \citep[see, e.g.,][for a case study of such process in the Fornax dwarf]{Zick.etal.2018}. 
Large scatter in $M_\star/L_{\rm UV}$ will boost the abundance of UV-bright galaxies and may help explain an overabundance of the UV-bright galaxies at $z\gtrsim 10$ observed by JWST \citep[e.g.,][]{Finkelstein.etal.2023,Wilkins.etal.2023,Yung.etal.2023}. 

%----------------------------------------------------------------------------------------------------------------------------
\subsection{From pre-disk Aurora stage to disk formation: qualitative changes in Galaxy's evolution}
%----------------------------------------------------------------------------------------------------------------------------

This work adds new details to the emerging picture of the rapid and profound transformation that the Milky Way underwent in its youth. Below we list the main changes in the properties of the Galaxy around the high redshift epoch corresponding to metallicities $\rm [Fe/H]\approx-1$.

{\it Spin-up.} The Milky Way was not born a disk. As we demonstrated in BK22, the earliest state of the Galaxy accessible to scrutiny through an archaeological record of its stars is that of i) high velocity dispersion and ii) little net rotation. Around $\rm [Fe/H]\approx-1.5$, the velocity ellipsoid of the in-situ stars is quasi-isotropic with a dispersion of ${\sim90}$ km/s. At these metallicities, there is some evidence for a modest net spin characterized by a median $V_{\phi}\approx50$ km/s. The median $v_\phi$ then increases promptly with increasing metallicity, and by $\rm[Fe/H]\approx-0.9$, the Galaxy has an established coherently rotating disk with a median tangential velocity of $V_{\rm \phi}\approx150$ km/s. This is revealed in Figures 4 and 5 of BK22 (see also Figure 3 of \citealt{Conroy2022} and Figure 6 of \citealt{Rix2022}). Numerical simulations show that the record of the original kinematic state of the Galaxy remains largely unaltered until the present day notwithstanding the intervening merger history \citep[see, e.g., BK22;][]{Mccluskey2023}.

{\it Scatter in elemental abundances.} As its bulk  stellar kinematics is reshaped, Galaxy's chemical properties evolve as well. The {\it Aurora} pre-disk stellar population exhibits a large scatter in most chemical abundance ratios. The origin of this scatter has not been pinned down yet, but it is hypothesised that continuous gas accretion and bursty stat-formation may play a role. Figure 7 of BK22 shows that dispersions in abundance ratios of most elements shrink on average by a factor of ${\sim2}$ on crossing the $\rm [Fe/H]\approx-1$ threshold. Some elements, such as nitrogen, exhibit a more dramatic decrease in abundance scatter, i.e. by a factor of ${\sim4}$ as the Galaxy forms its disk.

{\it Transition from clustered to regular star formation.} In BK22 we showed that the elements that exhibit the largest scatter in the Aurora population are N, Al, O, and Si, i.e. all of the established markers of anomalous chemical evolution characteristic of stars in GCs. It is therefore surmised that massive GCs had an important role in the formation and evolution of the high-redshift Milky Way, imprinted in the properties of the Aurora stellar population. In this work, we explore further the contribution of GCs to the Galactic stellar field and show that the fraction of the stars form in bound clusters is at its highest in Aurora, i.e. at $\rm [Fe/H]<-1$, and drops by more than an order of magnitude at $\rm [Fe/H]>-1$, where the Galactic disk forms.

{\it The  $\alpha$-bump.} There is evidence that not only the structural properties of the Galactic star formation changed drastically with the creation of the disk, but also its overall efficiency. The first obvious signature of the likely star-formation burst taking place around $\rm [Fe/H]\approx-1$ is in the rise and fall of the median $\alpha$-abundance ratios of both in-situ field stars and in-situ GCs as a function of metallicity (see middle panel of Figure~\ref{fig:gc_aur_acc}), which was pointed out and explored in detail by \citet{Conroy2022}.

{\it Change of slope of the metallicity distribution.} An increase in the pace of star formation betrayed by the rise of [Mg/Fe] is naturally imprinted in the shape of the metallicity distribution function. As left panel of Figure~\ref{fig:no_feh_elz} shows, the MDFs of the in-situ and the accreted stars agree at $\rm [Fe/H]<-1.5$, where the slopes of both distributions are close to ${\rm d} \log(n)/{\rm d[Fe/H]}\approx1$ \citep[also see][for discussion]{Rix2022}. At $\rm [Fe/H]\approx-1$, the MDF of the in-situ population has a clear inflection point: its slope abruptly changes to ${\rm d} \log(n)/{\rm d[Fe/H]}\approx2.3$. In comparison, the MDF of the accreted stars (mainly GS/E) shows no features around $\rm [Fe/H]\approx-1$.

\section{Conclusions}
\label{sec:conc}

Over the last decade, a consensus has been reached that a small number of the Milky Way field stars exhibit a chemical abundance pattern characteristic of that found in Galactic globular clusters. In particular, excess nitrogen abundance has been used widely as an effective GC fingerprint. While the globular cluster origin of these nitrogen-rich field stars is likely beyond doubt, the birthplace of the clusters themselves has remained unclear. This work provides a fresh re-assessment of the origin of the field stars enriched in nitrogen. For the first time, the origin of the bulk of such stars -- the low-metallicity in-situ Aurora population -- is identified unambiguously. Below we summarize our main results and spell out their implications. 
\begin{itemize}
     \item[(i)] We start by defining the selection criteria that relate directly to the properties of the observed GC stellar populations. The GC stars not only show elevated levels of nitrogen but also exhibit reduced oxygen as well as increased aluminium abundances. Accordingly, we identify field stars with high [N/O] and [Al/Fe] ratios, namely with [N/O]$>0.55$ and [Al/Fe]$>-0.1$ (see Figure~\ref{fig:select}). Relying on [N/O] instead of [N/Fe] also helps us to run a comparison with measurements of gas-phase abundances in high-redshift galaxies, which are referenced to oxygen.\\
    
    \item[(ii)] At the basis of our analysis is a new classification of the in-situ and accreted field stars and GCs. We first use [Al/Fe] in the metallicity range $\rm [Fe/H]\in [-1.4,-1]$ to estimate the boundary between in-situ and accreted objects in the plane of total energy and vertical component of the angular momentum $(E, L_z)$, as the high-[Al/Fe] and low-[Al/Fe] stars have very different $(E,L_z)$ distributions with a small overlap (see Figure~\ref{fig:elz_aurora}). We approximate the boundary with a simple parametric form  (see Equation~\ref{eq:sel}), but we have also tested our results against the classification done with a ``Gradient Boosted Trees'' machine learning method  instead \citep{Friedman.2001}, using {\tt GradientBoostingClassifier} class from the Sci-kit Learn package, and found no significant difference in classification accuracy. \\
      
    \item[(iii)] We show that simply using the $(E, L_z)$ boundary informed by distributions of stars with distinct [Al/Fe] ratios results in two population of GCs with distinct properties (see Figure~\ref{fig:gc_aur_acc}). For example, the accreted GCs show systematically lower [Mg/Fe] ratios at fixed [Fe/H] than the in-situ objects,  in-situ and accreted clusters occupy different regions of the [Mg/Fe]-[Al/Fe] space. Note that in our in-situ/accreted classification (i) we do not separate MW bulge and disk components and ii) we see no evidence for an additional low-energy component sometimes referred to as the Low-Energy group, Kraken/Heracles and Koala \citep[see][]{Massari2019,Krijssen2020,Forbes2020,Horta_heracles}.\\

    \item[(iv)] We show that the distribution of the selected high-[N/O] stars in the $(E, L_z)$ plane matches that of the Aurora population with high-[Al/Fe] and $-1.4<\rm [Fe/H]<-1$  (see Figures~\ref{fig:no_feh_elz} and ~\ref{fig:frac_energy}). Furthermore, we show that the radial distribution of high-[N/O] stars is very similar to that of the Aurora population (see Figure~\ref{fig:ratio_rad}). This indicates that the majority of the high-[N/O] stars formed in situ. This is not surprising because, if these stars form in GCs, not only the MW is expected to produce a lot more GCs than any of the accreted dwarfs due to the strong correlation between the total GC mass and the mass of the host \citep[e.g.,][]{Hudson.etal.2014,Forbes.etal.2018}, the MW-born GCs are on average older and have more time to dissolve and disrupt. While the $(E, L_z)$ distributions of Aurora and high-[N/O] stars are very similar, a small fraction of field stars enriched in nitrogen can possibly be assigned to the accreted population, more precisely to the GS/E merger. \\

    \item[(v)] We show that distributions of high-[N/O] stars in metallicity, energy, $L_z$, and galactocentric distance is similar to those of the in-situ GCs at metallicities $\rm [Fe/H]\lesssim -1$ (see Figures~\ref{fig:feh_dist_ns_gc},~\ref{fig:dgc_nrich_dist}) and is different from the corresponding distributions of GCs classified as accreted. At   $\rm [Fe/H]> -1$ the distributions of high-[N/O] stars and GCs are different, which indicates that at later times corresponding to such metallicities these stars form via a different route. \\
    
    \item[(vi)] We estimate the high-[N/O] fraction of the total stellar mass at a given metallicity, $f_{\rm N/O}\rm([Fe/H])$, and show that this fraction decreases rapidly with increasing metallicity (Figure~\ref{fig:frac_feh2}) from $2\%<f_{\rm N/O}<4\%$ at $\rm [Fe/H] \approx-1.4$ to $f_{\rm N/O}\approx0.04\%$ at $\rm [Fe/H]>-0.9$. In contrast, even if all high-[N/O] stars in the overlapping region of $(E, L_z)$ are assigned to the GS/E, the high-[N/O] fraction in the progenitor dwarf galaxy is some 5 times lower at $\rm [Fe/H]<-1$, i.e. $f_{\rm N/O}\approx0.8\%$ and shows no obvious trend with metallicity. \\
    
    \item[(vii)] Given that the estimated $f_{\rm N/O}$ fraction depends sensitively on the [N/O] threshold adopted for the selection of these peculiar stars, we measure the fraction of high-[N/O] stars in surviving Galactic GCs and show that $f_{\rm N/O, cl}$ scales with cluster initial mass (see Figure~\ref{fig:gc_mass}). Combining the relation between $f_{\rm N/O, cl}$ and initial mass with a model of the initial mass function of a population of freshly-born GCs, we show that the observed in-situ $f_{\rm N/O}$ implies that up to $f_{\rm cl}\approx 50\%-70\%$ of the stellar mass being formed in bound clusters at $\rm [Fe/H]\lesssim -1.5$ in the high-redshift MW (see Figure~\ref{fig:frac_feh2}) and up to $\approx 4-15\%$ at $\rm [Fe/H]=-1$ (see S~\ref{sec:fcl}).\\

    \item[(viii)] These results show that star formation in bound star clusters was a significant mode of star formation in the early Milky Way. In this context, we show that low-metallicity ($\rm [Fe/H]<-1$) in-situ stellar particles in the FIRE-2 simulations of Milky Way-sized galaxies have distributions of galactocentric distance and age-metallicity relations quite similar to those of the in-situ MW globular clusters (see Figure~\ref{fig:dgc_insitu_dist}).\\

    \item[(ix)] We use the estimated mass fraction of star formation in bound clusters as a function of metallicity to estimate the stellar mass of the Aurora population (stars with $[\rm Fe/H]<-1$ formed in-situ) of  $M_{\rm Aur} = 5\pm 2 \times 10^8 (1-f_{\rm disrupt})^{-1}\, M_\odot$, where $f_{\rm disrupt}$ is a fraction of clusters disrupted by $z=0$ expected to be $\approx 0.5-0.7$ in the models of cluster evolution (see \S~\ref{sec:maurora} for details). This indicates that the mass of this population is comparable to that of the overall MW stellar halo. Given that the Aurora stars are more centrally concentrated than the accreted halo, this component is expected to dominate the spheroidal population in the inner $\approx 10$ kpc of the Galaxy. \\

    \item[(x)] We argue that if the MW evolution is typical of the early stages of galaxies, the high fraction of star formation in massive bound clusters may help explain the anomalous high [N/O] ratio observed in the $z\approx 10.6$ galaxy GN-z11 (see Section~\ref{sec:gnz11}) and the abundance of UV-bright galaxies at $z\gtrsim 10$ (see Section~\ref{sec:discussion}).

\end{itemize}

\section*{Acknowledgments}

We are grateful to Oleg Gnedin for useful discussions about the fraction of disrupted globular clusters at different metallicities. We also wish to thank Stephanie Monty, Mark Gieles, Danny Horta, Harley Katz, Jason Sanders and GyuChul Myeong for their comments that helped improve the quality of this manuscript. VB wishes to thank Andy Bunker and Chiaki Kobayashi for stimulating discussions of the GN-z11's chemistry. AK was supported by the National Science Foundation grants AST-1714658 and AST-1911111 and NASA ATP grant 80NSSC20K0512.
This research made use of data from the European Space Agency mission Gaia
(\url{http://www.cosmos.esa.int/gaia}), processed by the Gaia Data
Processing and Analysis Consortium (DPAC,
\url{http://www.cosmos.esa.int/web/gaia/dpac/consortium}). Funding for the
DPAC has been provided by national institutions, in particular, the
institutions participating in the Gaia Multilateral Agreement. 

This
paper made used of the Whole Sky Database (wsdb) created by Sergey
Koposov and maintained at the Institute of Astronomy, Cambridge with
financial support from the Science \& Technology Facilities Council
(STFC) and the European Research Council (ERC). We also used FIRE-2 simulation public data  \citep[][\url{http://flathub.flatironinstitute.org/fire}]{Wetzel.etal.2023}, which are part of the Feedback In Realistic Environments (FIRE) project, generated using the Gizmo code \citep{hopkins15} and the FIRE-2 physics model \citep{hopkins_etal18}. 
Analyses presented in this paper were greatly aided by the following free software packages: {\tt NumPy} \citep{NumPy}, {\tt SciPy} \citep{scipy}, {\tt Matplotlib} \citep{matplotlib}, and {\tt Scikit-learn} \citep{sklearn}. We have also used the Astrophysics Data Service (\href{http://adsabs.harvard.edu/abstract_service.html}{\tt ADS}) and \href{https://arxiv.org}{\tt arXiv} preprint repository extensively during this project and the writing of the paper.

\section*{Data Availability}

This study uses  \verb|allStarLite-dr17-synspec_rev1| and \verb|apogee_astroNN-DR17| catalogues publicly available at \url{https://www.sdss.org/dr17/irspec/spectro_data/}. The catalog of the MW globular clusters with distances used in this study is publicly available at
\url{https://people.smp.uq.edu.au/HolgerBaumgardt/globular/}.

\bibliography{references}

\begin{thebibliography}{}
\makeatletter
\relax
\def\mn@urlcharsother{\let\do\@makeother \do\$\do\&\do\#\do\^\do\_\do\%\do\~}
\def\mn@doi{\begingroup\mn@urlcharsother \@ifnextchar [ {\mn@doi@}
  {\mn@doi@[]}}
\def\mn@doi@[#1]#2{\def\@tempa{#1}\ifx\@tempa\@empty \href
  {http://dx.doi.org/#2} {doi:#2}\else \href {http://dx.doi.org/#2} {#1}\fi
  \endgroup}
\def\mn@eprint#1#2{\mn@eprint@#1:#2::\@nil}
\def\mn@eprint@arXiv#1{\href {http://arxiv.org/abs/#1} {{\tt arXiv:#1}}}
\def\mn@eprint@dblp#1{\href {http://dblp.uni-trier.de/rec/bibtex/#1.xml}
  {dblp:#1}}
\def\mn@eprint@#1:#2:#3:#4\@nil{\def\@tempa {#1}\def\@tempb {#2}\def\@tempc
  {#3}\ifx \@tempc \@empty \let \@tempc \@tempb \let \@tempb \@tempa \fi \ifx
  \@tempb \@empty \def\@tempb {arXiv}\fi \@ifundefined
  {mn@eprint@\@tempb}{\@tempb:\@tempc}{\expandafter \expandafter \csname
  mn@eprint@\@tempb\endcsname \expandafter{\@tempc}}}

\bibitem[\protect\citeauthoryear{{Abdurro'uf} et~al.,}{{Abdurro'uf}
  et~al.}{2021}]{apogeedr17}
{Abdurro'uf} et~al., 2021, arXiv e-prints, \href
  {https://ui.adsabs.harvard.edu/abs/2021arXiv211202026A} {p. arXiv:2112.02026}

\bibitem[\protect\citeauthoryear{{Adamo} et~al.,}{{Adamo}
  et~al.}{2020a}]{Adamo.etal.2020rev}
{Adamo} A.,  et~al., 2020a, \mn@doi [\ssr] {10.1007/s11214-020-00690-x}, \href
  {https://ui.adsabs.harvard.edu/abs/2020SSRv..216...69A} {216, 69}

\bibitem[\protect\citeauthoryear{{Adamo} et~al.,}{{Adamo}
  et~al.}{2020b}]{Adamo.etal.2020}
{Adamo} A.,  et~al., 2020b, \mn@doi [\mnras] {10.1093/mnras/staa2380}, \href
  {https://ui.adsabs.harvard.edu/abs/2020MNRAS.499.3267A} {499, 3267}

\bibitem[\protect\citeauthoryear{{Arentsen} et~al.,}{{Arentsen}
  et~al.}{2020a}]{PIGS_I}
{Arentsen} A.,  et~al., 2020a, \mn@doi [\mnras] {10.1093/mnrasl/slz156}, \href
  {https://ui.adsabs.harvard.edu/abs/2020MNRAS.491L..11A} {491, L11}

\bibitem[\protect\citeauthoryear{{Arentsen} et~al.,}{{Arentsen}
  et~al.}{2020b}]{PIGS_II}
{Arentsen} A.,  et~al., 2020b, \mn@doi [\mnras] {10.1093/mnras/staa1661}, \href
  {https://ui.adsabs.harvard.edu/abs/2020MNRAS.496.4964A} {496, 4964}

\bibitem[\protect\citeauthoryear{{Bastian} \& {Lardo}}{{Bastian} \&
  {Lardo}}{2018}]{Bastian_Lardo2018}
{Bastian} N.,  {Lardo} C.,  2018, \mn@doi [\araa]
  {10.1146/annurev-astro-081817-051839}, \href
  {https://ui.adsabs.harvard.edu/abs/2018ARA&A..56...83B} {56, 83}

\bibitem[\protect\citeauthoryear{{Bastian}, {Lamers}, {de Mink}, {Longmore},
  {Goodwin}  \& {Gieles}}{{Bastian} et~al.}{2013}]{Bastian2013}
{Bastian} N.,  {Lamers} H.~J.~G.~L.~M.,  {de Mink} S.~E.,  {Longmore} S.~N.,
  {Goodwin} S.~P.,   {Gieles} M.,  2013, \mn@doi [\mnras]
  {10.1093/mnras/stt1745}, \href
  {https://ui.adsabs.harvard.edu/abs/2013MNRAS.436.2398B} {436, 2398}

\bibitem[\protect\citeauthoryear{{Baumgardt} \& {Makino}}{{Baumgardt} \&
  {Makino}}{2003}]{Baumgardt2003}
{Baumgardt} H.,  {Makino} J.,  2003, \mn@doi [\mnras]
  {10.1046/j.1365-8711.2003.06286.x}, \href
  {https://ui.adsabs.harvard.edu/abs/2003MNRAS.340..227B} {340, 227}

\bibitem[\protect\citeauthoryear{{Baumgardt} \& {Vasiliev}}{{Baumgardt} \&
  {Vasiliev}}{2021a}]{Baumgardt2021}
{Baumgardt} H.,  {Vasiliev} E.,  2021a, \mn@doi [\mnras]
  {10.1093/mnras/stab1474}, \href
  {https://ui.adsabs.harvard.edu/abs/2021MNRAS.505.5957B} {505, 5957}

\bibitem[\protect\citeauthoryear{{Baumgardt} \& {Vasiliev}}{{Baumgardt} \&
  {Vasiliev}}{2021b}]{Baumgardt_Vasiliev2021}
{Baumgardt} H.,  {Vasiliev} E.,  2021b, \mn@doi [\mnras]
  {10.1093/mnras/stab1474}, \href
  {https://ui.adsabs.harvard.edu/abs/2021MNRAS.505.5957B} {505, 5957}

\bibitem[\protect\citeauthoryear{{Bekki}, {Yahagi}, {Nagashima}  \&
  {Forbes}}{{Bekki} et~al.}{2008}]{Bekki.etal.2008}
{Bekki} K.,  {Yahagi} H.,  {Nagashima} M.,   {Forbes} D.~A.,  2008, \mn@doi
  [\mnras] {10.1111/j.1365-2966.2008.13318.x}, \href
  {https://ui.adsabs.harvard.edu/abs/2008MNRAS.387.1131B} {387, 1131}

\bibitem[\protect\citeauthoryear{{Belokurov} \& {Kravtsov}}{{Belokurov} \&
  {Kravtsov}}{2022}]{Aurora}
{Belokurov} V.,  {Kravtsov} A.,  2022, \mn@doi [\mnras]
  {10.1093/mnras/stac1267}, \href
  {https://ui.adsabs.harvard.edu/abs/2022MNRAS.514..689B} {514, 689}

\bibitem[\protect\citeauthoryear{{Belokurov}, {Erkal}, {Evans}, {Koposov}  \&
  {Deason}}{{Belokurov} et~al.}{2018}]{Belokurov2018}
{Belokurov} V.,  {Erkal} D.,  {Evans} N.~W.,  {Koposov} S.~E.,   {Deason}
  A.~J.,  2018, \mn@doi [\mnras] {10.1093/mnras/sty982}, \href
  {https://ui.adsabs.harvard.edu/abs/2018MNRAS.478..611B} {478, 611}

\bibitem[\protect\citeauthoryear{{Belokurov}, {Sanders}, {Fattahi}, {Smith},
  {Deason}, {Evans}  \& {Grand}}{{Belokurov} et~al.}{2020}]{Splash}
{Belokurov} V.,  {Sanders} J.~L.,  {Fattahi} A.,  {Smith} M.~C.,  {Deason}
  A.~J.,  {Evans} N.~W.,   {Grand} R. J.~J.,  2020, \mn@doi [\mnras]
  {10.1093/mnras/staa876}, \href
  {https://ui.adsabs.harvard.edu/abs/2020MNRAS.494.3880B} {494, 3880}

\bibitem[\protect\citeauthoryear{{Belokurov}, {Vasiliev}, {Deason}, {Koposov},
  {Fattahi}, {Dillamore}, {Davies}  \& {Grand}}{{Belokurov}
  et~al.}{2023}]{wrinkles}
{Belokurov} V.,  {Vasiliev} E.,  {Deason} A.~J.,  {Koposov} S.~E.,  {Fattahi}
  A.,  {Dillamore} A.~M.,  {Davies} E.~Y.,   {Grand} R. J.~J.,  2023, \mn@doi
  [\mnras] {10.1093/mnras/stac3436}, \href
  {https://ui.adsabs.harvard.edu/abs/2023MNRAS.518.6200B} {518, 6200}

\bibitem[\protect\citeauthoryear{{Bouwens}, {Illingworth}, {Oesch}, {Stefanon},
  {Naidu}, {van Leeuwen}  \& {Magee}}{{Bouwens} et~al.}{2023}]{Bowens2023}
{Bouwens} R.,  {Illingworth} G.,  {Oesch} P.,  {Stefanon} M.,  {Naidu} R.,
  {van Leeuwen} I.,   {Magee} D.,  2023, \mn@doi [\mnras]
  {10.1093/mnras/stad1014}, \href
  {https://ui.adsabs.harvard.edu/abs/2023MNRAS.tmp.1019B} {}

\bibitem[\protect\citeauthoryear{{Boylan-Kolchin}}{{Boylan-Kolchin}}{2018}]{Boylan.Kolchin.2018}
{Boylan-Kolchin} M.,  2018, \mn@doi [\mnras] {10.1093/mnras/sty1490}, \href
  {https://ui.adsabs.harvard.edu/abs/2018MNRAS.479..332B} {479, 332}

\bibitem[\protect\citeauthoryear{{Bunker} et~al.,}{{Bunker}
  et~al.}{2023}]{Bunker2023}
{Bunker} A.~J.,  et~al., 2023, \mn@doi [arXiv e-prints]
  {10.48550/arXiv.2302.07256}, \href
  {https://ui.adsabs.harvard.edu/abs/2023arXiv230207256B} {p. arXiv:2302.07256}

\bibitem[\protect\citeauthoryear{{Callingham}, {Cautun}, {Deason}, {Frenk},
  {Grand}  \& {Marinacci}}{{Callingham} et~al.}{2022}]{Callingham2022}
{Callingham} T.~M.,  {Cautun} M.,  {Deason} A.~J.,  {Frenk} C.~S.,  {Grand} R.
  J.~J.,   {Marinacci} F.,  2022, \mn@doi [\mnras] {10.1093/mnras/stac1145},
  \href {https://ui.adsabs.harvard.edu/abs/2022MNRAS.513.4107C} {513, 4107}

\bibitem[\protect\citeauthoryear{{Cameron}, {Katz}, {Rey}  \&
  {Saxena}}{{Cameron} et~al.}{2023}]{Cameron2023}
{Cameron} A.~J.,  {Katz} H.,  {Rey} M.~P.,   {Saxena} A.,  2023, \mn@doi [arXiv
  e-prints] {10.48550/arXiv.2302.10142}, \href
  {https://ui.adsabs.harvard.edu/abs/2023arXiv230210142C} {p. arXiv:2302.10142}

\bibitem[\protect\citeauthoryear{{Carollo}, {Martell}, {Beers}  \&
  {Freeman}}{{Carollo} et~al.}{2013}]{Carollo2013}
{Carollo} D.,  {Martell} S.~L.,  {Beers} T.~C.,   {Freeman} K.~C.,  2013,
  \mn@doi [\apj] {10.1088/0004-637X/769/2/87}, \href
  {https://ui.adsabs.harvard.edu/abs/2013ApJ...769...87C} {769, 87}

\bibitem[\protect\citeauthoryear{{Carretta} \& {Bragaglia}}{{Carretta} \&
  {Bragaglia}}{2022}]{Carretta2022}
{Carretta} E.,  {Bragaglia} A.,  2022, \mn@doi [\aap]
  {10.1051/0004-6361/202243211}, \href
  {https://ui.adsabs.harvard.edu/abs/2022A&A...660L...1C} {660, L1}

\bibitem[\protect\citeauthoryear{{Charbonnel}, {Schaerer}, {Prantzos},
  {Ram{\'\i}rez-Galeano}, {Fragos}, {Kuruvandothi}, {Marques-Chaves}  \&
  {Gieles}}{{Charbonnel} et~al.}{2023}]{Charbonnel2023}
{Charbonnel} C.,  {Schaerer} D.,  {Prantzos} N.,  {Ram{\'\i}rez-Galeano} L.,
  {Fragos} T.,  {Kuruvandothi} A.,  {Marques-Chaves} R.,   {Gieles} M.,  2023,
  \mn@doi [arXiv e-prints] {10.48550/arXiv.2303.07955}, \href
  {https://ui.adsabs.harvard.edu/abs/2023arXiv230307955C} {p. arXiv:2303.07955}

\bibitem[\protect\citeauthoryear{{Chen} \& {Gnedin}}{{Chen} \&
  {Gnedin}}{2023}]{Chen_Gnedin2023}
{Chen} Y.,  {Gnedin} O.~Y.,  2023, \mn@doi [\mnras] {10.1093/mnras/stad1328},
  \href {https://ui.adsabs.harvard.edu/abs/2023MNRAS.tmp.1307C} {}

\bibitem[\protect\citeauthoryear{{Choksi} \& {Gnedin}}{{Choksi} \&
  {Gnedin}}{2019}]{Choksi2019}
{Choksi} N.,  {Gnedin} O.~Y.,  2019, \mn@doi [\mnras] {10.1093/mnras/stz811},
  \href {https://ui.adsabs.harvard.edu/abs/2019MNRAS.486..331C} {486, 331}

\bibitem[\protect\citeauthoryear{{Choksi}, {Gnedin}  \& {Li}}{{Choksi}
  et~al.}{2018}]{Choksi2018}
{Choksi} N.,  {Gnedin} O.~Y.,   {Li} H.,  2018, \mn@doi [\mnras]
  {10.1093/mnras/sty1952}, \href
  {https://ui.adsabs.harvard.edu/abs/2018MNRAS.480.2343C} {480, 2343}

\bibitem[\protect\citeauthoryear{{Chon}, {Omukai}  \& {Schneider}}{{Chon}
  et~al.}{2021}]{Chon.etal.2021}
{Chon} S.,  {Omukai} K.,   {Schneider} R.,  2021, \mn@doi [\mnras]
  {10.1093/mnras/stab2497}, \href
  {https://ui.adsabs.harvard.edu/abs/2021MNRAS.508.4175C} {508, 4175}

\bibitem[\protect\citeauthoryear{{Conroy}}{{Conroy}}{2012}]{Conroy2012}
{Conroy} C.,  2012, \mn@doi [\apj] {10.1088/0004-637X/758/1/21}, \href
  {https://ui.adsabs.harvard.edu/abs/2012ApJ...758...21C} {758, 21}

\bibitem[\protect\citeauthoryear{{Conroy} et~al.,}{{Conroy}
  et~al.}{2022}]{Conroy2022}
{Conroy} C.,  et~al., 2022, \mn@doi [arXiv e-prints]
  {10.48550/arXiv.2204.02989}, \href
  {https://ui.adsabs.harvard.edu/abs/2022arXiv220402989C} {p. arXiv:2204.02989}

\bibitem[\protect\citeauthoryear{{Crowther}}{{Crowther}}{2007}]{Crowther.2007}
{Crowther} P.~A.,  2007, \mn@doi [\araa]
  {10.1146/annurev.astro.45.051806.110615}, \href
  {https://ui.adsabs.harvard.edu/abs/2007ARA&A..45..177C} {45, 177}

\bibitem[\protect\citeauthoryear{{Curtis-Lake} et~al.,}{{Curtis-Lake}
  et~al.}{2023}]{Curtis-Lake2023}
{Curtis-Lake} E.,  et~al., 2023, \mn@doi [Nature Astronomy]
  {10.1038/s41550-023-01918-w}, \href
  {https://ui.adsabs.harvard.edu/abs/2023NatAs.tmp...66C} {}

\bibitem[\protect\citeauthoryear{{D'Ercole}, {Vesperini}, {D'Antona},
  {McMillan}  \& {Recchi}}{{D'Ercole} et~al.}{2008}]{Dercole2008}
{D'Ercole} A.,  {Vesperini} E.,  {D'Antona} F.,  {McMillan} S. L.~W.,
  {Recchi} S.,  2008, \mn@doi [\mnras] {10.1111/j.1365-2966.2008.13915.x},
  \href {https://ui.adsabs.harvard.edu/abs/2008MNRAS.391..825D} {391, 825}

\bibitem[\protect\citeauthoryear{{Das}, {Hawkins}  \& {Jofr{\'e}}}{{Das}
  et~al.}{2020}]{Das2020}
{Das} P.,  {Hawkins} K.,   {Jofr{\'e}} P.,  2020, \mn@doi [\mnras]
  {10.1093/mnras/stz3537}, \href
  {https://ui.adsabs.harvard.edu/abs/2020MNRAS.493.5195D} {493, 5195}

\bibitem[\protect\citeauthoryear{{Deason}, {Belokurov}, {Evans}  \&
  {Johnston}}{{Deason} et~al.}{2013}]{Deason2013}
{Deason} A.~J.,  {Belokurov} V.,  {Evans} N.~W.,   {Johnston} K.~V.,  2013,
  \mn@doi [\apj] {10.1088/0004-637X/763/2/113}, \href
  {https://ui.adsabs.harvard.edu/abs/2013ApJ...763..113D} {763, 113}

\bibitem[\protect\citeauthoryear{{Deason}, {Belokurov}  \& {Sanders}}{{Deason}
  et~al.}{2019}]{Deason.etal.2019}
{Deason} A.~J.,  {Belokurov} V.,   {Sanders} J.~L.,  2019, \mn@doi [\mnras]
  {10.1093/mnras/stz2793}, \href
  {https://ui.adsabs.harvard.edu/abs/2019MNRAS.490.3426D} {490, 3426}

\bibitem[\protect\citeauthoryear{{Denissenkov} \& {Hartwick}}{{Denissenkov} \&
  {Hartwick}}{2014}]{Denissenkov.Hartwick.2014}
{Denissenkov} P.~A.,  {Hartwick} F.~D.~A.,  2014, \mn@doi [\mnras]
  {10.1093/mnrasl/slt133}, \href
  {https://ui.adsabs.harvard.edu/abs/2014MNRAS.437L..21D} {437, L21}

\bibitem[\protect\citeauthoryear{{Dib}, {Kim}  \& {Shadmehri}}{{Dib}
  et~al.}{2007}]{Dib.etal.2007}
{Dib} S.,  {Kim} J.,   {Shadmehri} M.,  2007, \mn@doi [\mnras]
  {10.1111/j.1745-3933.2007.00362.x}, \href
  {https://ui.adsabs.harvard.edu/abs/2007MNRAS.381L..40D} {381, L40}

\bibitem[\protect\citeauthoryear{{Dillamore}, {Belokurov}, {Font}  \&
  {McCarthy}}{{Dillamore} et~al.}{2022}]{Dillamore.etal.2022}
{Dillamore} A.~M.,  {Belokurov} V.,  {Font} A.~S.,   {McCarthy} I.~G.,  2022,
  \mn@doi [\mnras] {10.1093/mnras/stac1038}, \href
  {https://ui.adsabs.harvard.edu/abs/2022MNRAS.513.1867D} {513, 1867}

\bibitem[\protect\citeauthoryear{{Donlon} \& {Newberg}}{{Donlon} \&
  {Newberg}}{2023}]{Donlon2023}
{Donlon} T.,  {Newberg} H.~J.,  2023, \mn@doi [\apj]
  {10.3847/1538-4357/acb150}, \href
  {https://ui.adsabs.harvard.edu/abs/2023ApJ...944..169D} {944, 169}

\bibitem[\protect\citeauthoryear{{Donnan} et~al.,}{{Donnan}
  et~al.}{2023}]{Donnan2023}
{Donnan} C.~T.,  et~al., 2023, \mn@doi [\mnras] {10.1093/mnras/stac3472}, \href
  {https://ui.adsabs.harvard.edu/abs/2023MNRAS.518.6011D} {518, 6011}

\bibitem[\protect\citeauthoryear{{Dornan} \& {Harris}}{{Dornan} \&
  {Harris}}{2023}]{Dornan.Harris.2023}
{Dornan} V.,  {Harris} W.~E.,  2023, \mn@doi [arXiv e-prints]
  {10.48550/arXiv.2304.11210}, \href
  {https://ui.adsabs.harvard.edu/abs/2023arXiv230411210D} {p. arXiv:2304.11210}

\bibitem[\protect\citeauthoryear{{El-Badry}, {Quataert}, {Weisz}, {Choksi}  \&
  {Boylan-Kolchin}}{{El-Badry} et~al.}{2019}]{El_Badry_etal.2019}
{El-Badry} K.,  {Quataert} E.,  {Weisz} D.~R.,  {Choksi} N.,   {Boylan-Kolchin}
  M.,  2019, \mn@doi [\mnras] {10.1093/mnras/sty3007}, \href
  {https://ui.adsabs.harvard.edu/abs/2019MNRAS.482.4528E} {482, 4528}

\bibitem[\protect\citeauthoryear{{Fattahi} et~al.,}{{Fattahi}
  et~al.}{2019}]{Fattahi.etal.2019}
{Fattahi} A.,  et~al., 2019, \mn@doi [\mnras] {10.1093/mnras/stz159}, \href
  {https://ui.adsabs.harvard.edu/abs/2019MNRAS.484.4471F} {484, 4471}

\bibitem[\protect\citeauthoryear{{Fern{\'a}ndez-Trincado}
  et~al.,}{{Fern{\'a}ndez-Trincado} et~al.}{2017}]{Fernandez_Trincado2017}
{Fern{\'a}ndez-Trincado} J.~G.,  et~al., 2017, \mn@doi [\apjl]
  {10.3847/2041-8213/aa8032}, \href
  {https://ui.adsabs.harvard.edu/abs/2017ApJ...846L...2F} {846, L2}

\bibitem[\protect\citeauthoryear{{Fern{\'a}ndez-Trincado}, {Chaves-Velasquez},
  {P{\'e}rez-Villegas}, {Vieira}, {Moreno}, {Ortigoza-Urdaneta}  \&
  {Vega-Neme}}{{Fern{\'a}ndez-Trincado} et~al.}{2020a}]{Trincado2020c}
{Fern{\'a}ndez-Trincado} J.~G.,  {Chaves-Velasquez} L.,  {P{\'e}rez-Villegas}
  A.,  {Vieira} K.,  {Moreno} E.,  {Ortigoza-Urdaneta} M.,   {Vega-Neme} L.,
  2020a, \mn@doi [\mnras] {10.1093/mnras/staa1386}, \href
  {https://ui.adsabs.harvard.edu/abs/2020MNRAS.495.4113F} {495, 4113}

\bibitem[\protect\citeauthoryear{{Fern{\'a}ndez-Trincado}, {Beers}, {Minniti},
  {Tang}, {Villanova}, {Geisler}, {P{\'e}rez-Villegas}  \&
  {Vieira}}{{Fern{\'a}ndez-Trincado} et~al.}{2020b}]{Fernandez_Trincado2020b}
{Fern{\'a}ndez-Trincado} J.~G.,  {Beers} T.~C.,  {Minniti} D.,  {Tang} B.,
  {Villanova} S.,  {Geisler} D.,  {P{\'e}rez-Villegas} A.,   {Vieira} K.,
  2020b, \mn@doi [\aap] {10.1051/0004-6361/202039207}, \href
  {https://ui.adsabs.harvard.edu/abs/2020A&A...643L...4F} {643, L4}

\bibitem[\protect\citeauthoryear{{Fern{\'a}ndez-Trincado}, {Beers}  \&
  {Minniti}}{{Fern{\'a}ndez-Trincado} et~al.}{2020c}]{Jurassic}
{Fern{\'a}ndez-Trincado} J.~G.,  {Beers} T.~C.,   {Minniti} D.,  2020c, \mn@doi
  [\aap] {10.1051/0004-6361/202039434}, \href
  {https://ui.adsabs.harvard.edu/abs/2020A&A...644A..83F} {644, A83}

\bibitem[\protect\citeauthoryear{{Fern{\'a}ndez-Trincado}
  et~al.,}{{Fern{\'a}ndez-Trincado} et~al.}{2020d}]{Fernandez_Trincado2020}
{Fern{\'a}ndez-Trincado} J.~G.,  et~al., 2020d, \mn@doi [\apjl]
  {10.3847/2041-8213/abc01d}, \href
  {https://ui.adsabs.harvard.edu/abs/2020ApJ...903L..17F} {903, L17}

\bibitem[\protect\citeauthoryear{{Fern{\'a}ndez-Trincado}
  et~al.,}{{Fern{\'a}ndez-Trincado} et~al.}{2022}]{Fernandez_Trincado2022}
{Fern{\'a}ndez-Trincado} J.~G.,  et~al., 2022, \mn@doi [\aap]
  {10.1051/0004-6361/202243195}, \href
  {https://ui.adsabs.harvard.edu/abs/2022A&A...663A.126F} {663, A126}

\bibitem[\protect\citeauthoryear{{Feuillet}, {Feltzing}, {Sahlholdt}  \&
  {Casagrande}}{{Feuillet} et~al.}{2020}]{Feuillet2020}
{Feuillet} D.~K.,  {Feltzing} S.,  {Sahlholdt} C.~L.,   {Casagrande} L.,  2020,
  \mn@doi [\mnras] {10.1093/mnras/staa1888}, \href
  {https://ui.adsabs.harvard.edu/abs/2020MNRAS.497..109F} {497, 109}

\bibitem[\protect\citeauthoryear{{Finkelstein} et~al.,}{{Finkelstein}
  et~al.}{2023a}]{Finkelstein2023}
{Finkelstein} S.~L.,  et~al., 2023a, \mn@doi [\apjl]
  {10.3847/2041-8213/acade4}, \href
  {https://ui.adsabs.harvard.edu/abs/2023ApJ...946L..13F} {946, L13}

\bibitem[\protect\citeauthoryear{{Finkelstein} et~al.,}{{Finkelstein}
  et~al.}{2023b}]{Finkelstein.etal.2023}
{Finkelstein} S.~L.,  et~al., 2023b, \mn@doi [\apjl]
  {10.3847/2041-8213/acade4}, \href
  {https://ui.adsabs.harvard.edu/abs/2023ApJ...946L..13F} {946, L13}

\bibitem[\protect\citeauthoryear{{Forbes}}{{Forbes}}{2020}]{Forbes2020}
{Forbes} D.~A.,  2020, \mn@doi [\mnras] {10.1093/mnras/staa245}, \href
  {https://ui.adsabs.harvard.edu/abs/2020MNRAS.493..847F} {493, 847}

\bibitem[\protect\citeauthoryear{{Forbes}, {Read}, {Gieles}  \&
  {Collins}}{{Forbes} et~al.}{2018}]{Forbes.etal.2018}
{Forbes} D.~A.,  {Read} J.~I.,  {Gieles} M.,   {Collins} M. L.~M.,  2018,
  \mn@doi [\mnras] {10.1093/mnras/sty2584}, \href
  {https://ui.adsabs.harvard.edu/abs/2018MNRAS.481.5592F} {481, 5592}

\bibitem[\protect\citeauthoryear{Friedman}{Friedman}{2001}]{Friedman.2001}
Friedman J.~H.,  2001, \mn@doi [The Annals of Statistics]
  {10.1214/aos/1013203451}, 29, 1189

\bibitem[\protect\citeauthoryear{{Gaia Collaboration} et~al.,}{{Gaia
  Collaboration} et~al.}{2021}]{gaia_edr3}
{Gaia Collaboration} et~al., 2021, \mn@doi [\aap]
  {10.1051/0004-6361/202039657}, \href
  {https://ui.adsabs.harvard.edu/abs/2021A&A...649A...1G} {649, A1}

\bibitem[\protect\citeauthoryear{{Gaia Collaboration} et~al.,}{{Gaia
  Collaboration} et~al.}{2022}]{Drimmel2022}
{Gaia Collaboration} et~al., 2022, arXiv e-prints, \href
  {https://ui.adsabs.harvard.edu/abs/2022arXiv220606207G} {p. arXiv:2206.06207}

\bibitem[\protect\citeauthoryear{{Gieles} \& {Gnedin}}{{Gieles} \&
  {Gnedin}}{2023}]{Gieles2023}
{Gieles} M.,  {Gnedin} O.,  2023, \mn@doi [arXiv e-prints]
  {10.48550/arXiv.2303.03791}, \href
  {https://ui.adsabs.harvard.edu/abs/2023arXiv230303791G} {p. arXiv:2303.03791}

\bibitem[\protect\citeauthoryear{{Gratton}, {Bragaglia}, {Carretta}, {D'Orazi},
  {Lucatello}  \& {Sollima}}{{Gratton} et~al.}{2019}]{Gratton2019}
{Gratton} R.,  {Bragaglia} A.,  {Carretta} E.,  {D'Orazi} V.,  {Lucatello} S.,
   {Sollima} A.,  2019, \mn@doi [\aapr] {10.1007/s00159-019-0119-3}, \href
  {https://ui.adsabs.harvard.edu/abs/2019A&ARv..27....8G} {27, 8}

\bibitem[\protect\citeauthoryear{{Gravity Collaboration} et~al.,}{{Gravity
  Collaboration} et~al.}{2022}]{GRAVITY2022}
{Gravity Collaboration} et~al., 2022, \mn@doi [\aap]
  {10.1051/0004-6361/202142465}, \href
  {https://ui.adsabs.harvard.edu/abs/2022A&A...657L..12G} {657, L12}

\bibitem[\protect\citeauthoryear{{Harikane} et~al.,}{{Harikane}
  et~al.}{2023}]{Harikane2023}
{Harikane} Y.,  et~al., 2023, \mn@doi [\apjs] {10.3847/1538-4365/acaaa9}, \href
  {https://ui.adsabs.harvard.edu/abs/2023ApJS..265....5H} {265, 5}

\bibitem[\protect\citeauthoryear{{Harris}}{{Harris}}{2010}]{Harris2010}
{Harris} W.~E.,  2010, \mn@doi [arXiv e-prints] {10.48550/arXiv.1012.3224},
  \href {https://ui.adsabs.harvard.edu/abs/2010arXiv1012.3224H} {p.
  arXiv:1012.3224}

\bibitem[\protect\citeauthoryear{{Harris}, {Blakeslee}  \& {Harris}}{{Harris}
  et~al.}{2017}]{Harris.etal.2017}
{Harris} W.~E.,  {Blakeslee} J.~P.,   {Harris} G. L.~H.,  2017, \mn@doi [\apj]
  {10.3847/1538-4357/836/1/67}, \href
  {https://ui.adsabs.harvard.edu/abs/2017ApJ...836...67H} {836, 67}

\bibitem[\protect\citeauthoryear{{Hasselquist} et~al.,}{{Hasselquist}
  et~al.}{2021}]{Hasselquist2021}
{Hasselquist} S.,  et~al., 2021, \mn@doi [\apj] {10.3847/1538-4357/ac25f9},
  \href {https://ui.adsabs.harvard.edu/abs/2021ApJ...923..172H} {923, 172}

\bibitem[\protect\citeauthoryear{{Hawkins}, {Jofr{\'e}}, {Masseron}  \&
  {Gilmore}}{{Hawkins} et~al.}{2015}]{Hawkins2015}
{Hawkins} K.,  {Jofr{\'e}} P.,  {Masseron} T.,   {Gilmore} G.,  2015, \mn@doi
  [\mnras] {10.1093/mnras/stv1586}, \href
  {https://ui.adsabs.harvard.edu/abs/2015MNRAS.453..758H} {453, 758}

\bibitem[\protect\citeauthoryear{{Helmi}, {Babusiaux}, {Koppelman}, {Massari},
  {Veljanoski}  \& {Brown}}{{Helmi} et~al.}{2018}]{Helmi2018}
{Helmi} A.,  {Babusiaux} C.,  {Koppelman} H.~H.,  {Massari} D.,  {Veljanoski}
  J.,   {Brown} A. G.~A.,  2018, \mn@doi [\nat] {10.1038/s41586-018-0625-x},
  \href {https://ui.adsabs.harvard.edu/abs/2018Natur.563...85H} {563, 85}

\bibitem[\protect\citeauthoryear{{Hopkins}}{{Hopkins}}{2015}]{hopkins15}
{Hopkins} P.~F.,  2015, \mn@doi [\mnras] {10.1093/mnras/stv195}, \href
  {https://ui.adsabs.harvard.edu/abs/2015MNRAS.450...53H} {450, 53}

\bibitem[\protect\citeauthoryear{{Hopkins} et~al.,}{{Hopkins}
  et~al.}{2018a}]{Hopkins.etal.2018}
{Hopkins} P.~F.,  et~al., 2018a, \mn@doi [\mnras] {10.1093/mnras/sty1690},
  \href {https://ui.adsabs.harvard.edu/abs/2018MNRAS.480..800H} {480, 800}

\bibitem[\protect\citeauthoryear{{Hopkins} et~al.,}{{Hopkins}
  et~al.}{2018b}]{hopkins_etal18}
{Hopkins} P.~F.,  et~al., 2018b, \mn@doi [\mnras] {10.1093/mnras/sty1690},
  \href {https://ui.adsabs.harvard.edu/abs/2018MNRAS.480..800H} {480, 800}

\bibitem[\protect\citeauthoryear{{Horta} et~al.,}{{Horta}
  et~al.}{2020}]{Horta_GC}
{Horta} D.,  et~al., 2020, \mn@doi [\mnras] {10.1093/mnras/staa478}, \href
  {https://ui.adsabs.harvard.edu/abs/2020MNRAS.493.3363H} {493, 3363}

\bibitem[\protect\citeauthoryear{{Horta} et~al.,}{{Horta}
  et~al.}{2021a}]{Horta_heracles}
{Horta} D.,  et~al., 2021a, \mn@doi [\mnras] {10.1093/mnras/staa2987}, \href
  {https://ui.adsabs.harvard.edu/abs/2021MNRAS.500.1385H} {500, 1385}

\bibitem[\protect\citeauthoryear{{Horta} et~al.,}{{Horta}
  et~al.}{2021b}]{Horta2021}
{Horta} D.,  et~al., 2021b, \mn@doi [\mnras] {10.1093/mnras/staa3598}, \href
  {https://ui.adsabs.harvard.edu/abs/2021MNRAS.500.5462H} {500, 5462}

\bibitem[\protect\citeauthoryear{{Horta} et~al.,}{{Horta}
  et~al.}{2023a}]{Horta2023}
{Horta} D.,  et~al., 2023a, \mn@doi [\mnras] {10.1093/mnras/stac3179}, \href
  {https://ui.adsabs.harvard.edu/abs/2023MNRAS.520.5671H} {520, 5671}

\bibitem[\protect\citeauthoryear{{Horta} et~al.,}{{Horta}
  et~al.}{2023b}]{Horta_FIRE}
{Horta} D.,  et~al., 2023b, \mn@doi [\apj] {10.3847/1538-4357/acae87}, \href
  {https://ui.adsabs.harvard.edu/abs/2023ApJ...943..158H} {943, 158}

\bibitem[\protect\citeauthoryear{{Hudson}, {Harris}  \& {Harris}}{{Hudson}
  et~al.}{2014}]{Hudson.etal.2014}
{Hudson} M.~J.,  {Harris} G.~L.,   {Harris} W.~E.,  2014, \mn@doi [\apjl]
  {10.1088/2041-8205/787/1/L5}, \href
  {https://ui.adsabs.harvard.edu/abs/2014ApJ...787L...5H} {787, L5}

\bibitem[\protect\citeauthoryear{Hunter}{Hunter}{2007}]{matplotlib}
Hunter J.~D.,  2007, \mn@doi [Computing In Science \& Engineering]
  {10.1109/MCSE.2007.55}, 9, 90

\bibitem[\protect\citeauthoryear{{Iorio} \& {Belokurov}}{{Iorio} \&
  {Belokurov}}{2021}]{Iorio2021}
{Iorio} G.,  {Belokurov} V.,  2021, \mn@doi [\mnras] {10.1093/mnras/stab005},
  \href {https://ui.adsabs.harvard.edu/abs/2021MNRAS.502.5686I} {502, 5686}

\bibitem[\protect\citeauthoryear{{Johnson} et~al.,}{{Johnson}
  et~al.}{2017}]{Johnson.etal.2017}
{Johnson} T.~L.,  et~al., 2017, \mn@doi [\apjl] {10.3847/2041-8213/aa7516},
  \href {https://ui.adsabs.harvard.edu/abs/2017ApJ...843L..21J} {843, L21}

\bibitem[\protect\citeauthoryear{{Johnson}, {Weinberg}, {Vincenzo}, {Bird}  \&
  {Griffith}}{{Johnson} et~al.}{2023}]{Johnson2023}
{Johnson} J.~W.,  {Weinberg} D.~H.,  {Vincenzo} F.,  {Bird} J.~C.,   {Griffith}
  E.~J.,  2023, \mn@doi [\mnras] {10.1093/mnras/stad057}, \href
  {https://ui.adsabs.harvard.edu/abs/2023MNRAS.520..782J} {520, 782}

\bibitem[\protect\citeauthoryear{Jones, Oliphant, Peterson  et~al.}{Jones
  et~al.}{01  }]{scipy}
Jones E.,  Oliphant T.,  Peterson P.,   et~al., 2001--, {SciPy}: Open source
  scientific tools for {Python}, \url {http://www.scipy.org/}

\bibitem[\protect\citeauthoryear{{Kim}, {Lee}, {Beers}  \& {Kim}}{{Kim}
  et~al.}{2023}]{Kim.etal.2023}
{Kim} C.,  {Lee} Y.~S.,  {Beers} T.~C.,   {Kim} Y.~K.,  2023, \mn@doi [arXiv
  e-prints] {10.48550/arXiv.2305.04025}, \href
  {https://ui.adsabs.harvard.edu/abs/2023arXiv230504025K} {p. arXiv:2305.04025}

\bibitem[\protect\citeauthoryear{{Kisku} et~al.,}{{Kisku}
  et~al.}{2021}]{Kisku2021}
{Kisku} S.,  et~al., 2021, \mn@doi [\mnras] {10.1093/mnras/stab525}, \href
  {https://ui.adsabs.harvard.edu/abs/2021MNRAS.504.1657K} {504, 1657}

\bibitem[\protect\citeauthoryear{{Kobayashi}, {Karakas}  \&
  {Lugaro}}{{Kobayashi} et~al.}{2020}]{Kobayashi2020}
{Kobayashi} C.,  {Karakas} A.~I.,   {Lugaro} M.,  2020, \mn@doi [\apj]
  {10.3847/1538-4357/abae65}, \href
  {https://ui.adsabs.harvard.edu/abs/2020ApJ...900..179K} {900, 179}

\bibitem[\protect\citeauthoryear{{Koch-Hansen}, {Hansen}  \&
  {McWilliam}}{{Koch-Hansen} et~al.}{2021}]{Koch-Hansen2021}
{Koch-Hansen} A.~J.,  {Hansen} C.~J.,   {McWilliam} A.,  2021, \mn@doi [\aap]
  {10.1051/0004-6361/202141130}, \href
  {https://ui.adsabs.harvard.edu/abs/2021A&A...653A...2K} {653, A2}

\bibitem[\protect\citeauthoryear{{Koch}, {Grebel}  \& {Martell}}{{Koch}
  et~al.}{2019}]{Koch2019}
{Koch} A.,  {Grebel} E.~K.,   {Martell} S.~L.,  2019, \mn@doi [\aap]
  {10.1051/0004-6361/201834825}, \href
  {https://ui.adsabs.harvard.edu/abs/2019A&A...625A..75K} {625, A75}

\bibitem[\protect\citeauthoryear{{Kravtsov} \& {Gnedin}}{{Kravtsov} \&
  {Gnedin}}{2005}]{Kravtsov2005}
{Kravtsov} A.~V.,  {Gnedin} O.~Y.,  2005, \mn@doi [\apj] {10.1086/428636},
  \href {https://ui.adsabs.harvard.edu/abs/2005ApJ...623..650K} {623, 650}

\bibitem[\protect\citeauthoryear{{Kravtsov}, {Vikhlinin}  \&
  {Meshcheryakov}}{{Kravtsov} et~al.}{2018}]{Kravtsov.etal.2018}
{Kravtsov} A.~V.,  {Vikhlinin} A.~A.,   {Meshcheryakov} A.~V.,  2018, \mn@doi
  [Astronomy Letters] {10.1134/S1063773717120015}, \href
  {https://ui.adsabs.harvard.edu/abs/2018AstL...44....8K} {44, 8}

\bibitem[\protect\citeauthoryear{{Kruijssen}}{{Kruijssen}}{2012}]{Kruijssen2012}
{Kruijssen} J.~M.~D.,  2012, \mn@doi [\mnras]
  {10.1111/j.1365-2966.2012.21923.x}, \href
  {https://ui.adsabs.harvard.edu/abs/2012MNRAS.426.3008K} {426, 3008}

\bibitem[\protect\citeauthoryear{{Kruijssen}, {Pfeffer}, {Reina-Campos},
  {Crain}  \& {Bastian}}{{Kruijssen} et~al.}{2019}]{Krujssen2019}
{Kruijssen} J.~M.~D.,  {Pfeffer} J.~L.,  {Reina-Campos} M.,  {Crain} R.~A.,
  {Bastian} N.,  2019, \mn@doi [\mnras] {10.1093/mnras/sty1609}, \href
  {https://ui.adsabs.harvard.edu/abs/2019MNRAS.486.3180K} {486, 3180}

\bibitem[\protect\citeauthoryear{{Kruijssen} et~al.,}{{Kruijssen}
  et~al.}{2020}]{Krijssen2020}
{Kruijssen} J.~M.~D.,  et~al., 2020, \mn@doi [\mnras] {10.1093/mnras/staa2452},
  \href {https://ui.adsabs.harvard.edu/abs/2020MNRAS.498.2472K} {498, 2472}

\bibitem[\protect\citeauthoryear{{Krumholz}, {McKee}  \&
  {Bland-Hawthorn}}{{Krumholz} et~al.}{2019}]{Krumholz2019}
{Krumholz} M.~R.,  {McKee} C.~F.,   {Bland-Hawthorn} J.,  2019, \mn@doi [\araa]
  {10.1146/annurev-astro-091918-104430}, \href
  {https://ui.adsabs.harvard.edu/abs/2019ARA&A..57..227K} {57, 227}

\bibitem[\protect\citeauthoryear{{Leung} \& {Bovy}}{{Leung} \&
  {Bovy}}{2019}]{Leung2019}
{Leung} H.~W.,  {Bovy} J.,  2019, \mn@doi [\mnras] {10.1093/mnras/stz2245},
  \href {https://ui.adsabs.harvard.edu/abs/2019MNRAS.489.2079L} {489, 2079}

\bibitem[\protect\citeauthoryear{{Lind} et~al.,}{{Lind}
  et~al.}{2015}]{Lind2015}
{Lind} K.,  et~al., 2015, \mn@doi [\aap] {10.1051/0004-6361/201425554}, \href
  {https://ui.adsabs.harvard.edu/abs/2015A&A...575L..12L} {575, L12}

\bibitem[\protect\citeauthoryear{{Lindegren} et~al.,}{{Lindegren}
  et~al.}{2021}]{Lindegren2021}
{Lindegren} L.,  et~al., 2021, \mn@doi [\aap] {10.1051/0004-6361/202039709},
  \href {https://ui.adsabs.harvard.edu/abs/2021A&A...649A...2L} {649, A2}

\bibitem[\protect\citeauthoryear{{Livermore} et~al.,}{{Livermore}
  et~al.}{2015}]{Livermore.etal.2015}
{Livermore} R.~C.,  et~al., 2015, \mn@doi [\mnras] {10.1093/mnras/stv686},
  \href {https://ui.adsabs.harvard.edu/abs/2015MNRAS.450.1812L} {450, 1812}

\bibitem[\protect\citeauthoryear{{Lodders}}{{Lodders}}{2019}]{Lodders2019}
{Lodders} K.,  2019, \mn@doi [arXiv e-prints] {10.48550/arXiv.1912.00844},
  \href {https://ui.adsabs.harvard.edu/abs/2019arXiv191200844L} {p.
  arXiv:1912.00844}

\bibitem[\protect\citeauthoryear{{Mackereth} \& {Bovy}}{{Mackereth} \&
  {Bovy}}{2018}]{Mackereth2018}
{Mackereth} J.~T.,  {Bovy} J.,  2018, \mn@doi [\pasp]
  {10.1088/1538-3873/aadcdd}, \href
  {https://ui.adsabs.harvard.edu/abs/2018PASP..130k4501M} {130, 114501}

\bibitem[\protect\citeauthoryear{{Mackereth} et~al.,}{{Mackereth}
  et~al.}{2019}]{Mackereth2019}
{Mackereth} J.~T.,  et~al., 2019, \mn@doi [\mnras] {10.1093/mnras/sty2955},
  \href {https://ui.adsabs.harvard.edu/abs/2019MNRAS.482.3426M} {482, 3426}

\bibitem[\protect\citeauthoryear{{Maeder} \& {Meynet}}{{Maeder} \&
  {Meynet}}{2006}]{Maeder.Meynet.2006}
{Maeder} A.,  {Meynet} G.,  2006, \mn@doi [\aap] {10.1051/0004-6361:200600012},
  \href {https://ui.adsabs.harvard.edu/abs/2006A&A...448L..37M} {448, L37}

\bibitem[\protect\citeauthoryear{{Martell} \& {Grebel}}{{Martell} \&
  {Grebel}}{2010}]{Martell_Grebel2010}
{Martell} S.~L.,  {Grebel} E.~K.,  2010, \mn@doi [\aap]
  {10.1051/0004-6361/201014135}, \href
  {https://ui.adsabs.harvard.edu/abs/2010A&A...519A..14M} {519, A14}

\bibitem[\protect\citeauthoryear{{Martell}, {Smolinski}, {Beers}  \&
  {Grebel}}{{Martell} et~al.}{2011}]{Martell2011}
{Martell} S.~L.,  {Smolinski} J.~P.,  {Beers} T.~C.,   {Grebel} E.~K.,  2011,
  \mn@doi [\aap] {10.1051/0004-6361/201117644}, \href
  {https://ui.adsabs.harvard.edu/abs/2011A&A...534A.136M} {534, A136}

\bibitem[\protect\citeauthoryear{{Martell} et~al.,}{{Martell}
  et~al.}{2016}]{Martell2016}
{Martell} S.~L.,  et~al., 2016, \mn@doi [\apj] {10.3847/0004-637X/825/2/146},
  \href {https://ui.adsabs.harvard.edu/abs/2016ApJ...825..146M} {825, 146}

\bibitem[\protect\citeauthoryear{{Massari}, {Koppelman}  \& {Helmi}}{{Massari}
  et~al.}{2019}]{Massari2019}
{Massari} D.,  {Koppelman} H.~H.,   {Helmi} A.,  2019, \mn@doi [\aap]
  {10.1051/0004-6361/201936135}, \href
  {https://ui.adsabs.harvard.edu/abs/2019A&A...630L...4M} {630, L4}

\bibitem[\protect\citeauthoryear{{McCluskey}, {Wetzel}, {Loebman}, {Moreno}  \&
  {Faucher-Giguere}}{{McCluskey} et~al.}{2023}]{Mccluskey2023}
{McCluskey} F.,  {Wetzel} A.,  {Loebman} S.~R.,  {Moreno} J.,
  {Faucher-Giguere} C.-A.,  2023, \mn@doi [arXiv e-prints]
  {10.48550/arXiv.2303.14210}, \href
  {https://ui.adsabs.harvard.edu/abs/2023arXiv230314210M} {p. arXiv:2303.14210}

\bibitem[\protect\citeauthoryear{{McKenzie} et~al.,}{{McKenzie}
  et~al.}{2022}]{McKenzie2022}
{McKenzie} M.,  et~al., 2022, \mn@doi [\mnras] {10.1093/mnras/stac2254}, \href
  {https://ui.adsabs.harvard.edu/abs/2022MNRAS.516.3515M} {516, 3515}

\bibitem[\protect\citeauthoryear{{Milone} \& {Marino}}{{Milone} \&
  {Marino}}{2022}]{Miloni_Marino2022}
{Milone} A.~P.,  {Marino} A.~F.,  2022, \mn@doi [Universe]
  {10.3390/universe8070359}, \href
  {https://ui.adsabs.harvard.edu/abs/2022Univ....8..359M} {8, 359}

\bibitem[\protect\citeauthoryear{{Milone} et~al.,}{{Milone}
  et~al.}{2017}]{Milone2017}
{Milone} A.~P.,  et~al., 2017, \mn@doi [\mnras] {10.1093/mnras/stw2531}, \href
  {https://ui.adsabs.harvard.edu/abs/2017MNRAS.464.3636M} {464, 3636}

\bibitem[\protect\citeauthoryear{{Milone} et~al.,}{{Milone}
  et~al.}{2020}]{Milone2020}
{Milone} A.~P.,  et~al., 2020, \mn@doi [\mnras] {10.1093/mnras/stz2999}, \href
  {https://ui.adsabs.harvard.edu/abs/2020MNRAS.491..515M} {491, 515}

\bibitem[\protect\citeauthoryear{{Monty} et~al.,}{{Monty}
  et~al.}{2023}]{Monty2023}
{Monty} S.,  et~al., 2023, \mn@doi [\mnras] {10.1093/mnras/stad1154}, \href
  {https://ui.adsabs.harvard.edu/abs/2023MNRAS.522.4404M} {522, 4404}

\bibitem[\protect\citeauthoryear{{Myeong}, {Evans}, {Belokurov}, {Sanders}  \&
  {Koposov}}{{Myeong} et~al.}{2018}]{Myeong2018}
{Myeong} G.~C.,  {Evans} N.~W.,  {Belokurov} V.,  {Sanders} J.~L.,   {Koposov}
  S.~E.,  2018, \mn@doi [\apjl] {10.3847/2041-8213/aad7f7}, \href
  {https://ui.adsabs.harvard.edu/abs/2018ApJ...863L..28M} {863, L28}

\bibitem[\protect\citeauthoryear{{Myeong}, {Vasiliev}, {Iorio}, {Evans}  \&
  {Belokurov}}{{Myeong} et~al.}{2019}]{Myeong2019}
{Myeong} G.~C.,  {Vasiliev} E.,  {Iorio} G.,  {Evans} N.~W.,   {Belokurov} V.,
  2019, \mn@doi [\mnras] {10.1093/mnras/stz1770}, \href
  {https://ui.adsabs.harvard.edu/abs/2019MNRAS.488.1235M} {488, 1235}

\bibitem[\protect\citeauthoryear{{Myeong}, {Belokurov}, {Aguado}, {Evans},
  {Caldwell}  \& {Bradley}}{{Myeong} et~al.}{2022}]{Myeong2022}
{Myeong} G.~C.,  {Belokurov} V.,  {Aguado} D.~S.,  {Evans} N.~W.,  {Caldwell}
  N.,   {Bradley} J.,  2022, \mn@doi [\apj] {10.3847/1538-4357/ac8d68}, \href
  {https://ui.adsabs.harvard.edu/abs/2022ApJ...938...21M} {938, 21}

\bibitem[\protect\citeauthoryear{{Naab} \& {Ostriker}}{{Naab} \&
  {Ostriker}}{2017}]{Naab.Ostriker.2017}
{Naab} T.,  {Ostriker} J.~P.,  2017, \mn@doi [\araa]
  {10.1146/annurev-astro-081913-040019}, \href
  {https://ui.adsabs.harvard.edu/abs/2017ARA&A..55...59N} {55, 59}

\bibitem[\protect\citeauthoryear{{Nadler} et~al.,}{{Nadler}
  et~al.}{2020}]{Nadler.etal.2020}
{Nadler} E.~O.,  et~al., 2020, \mn@doi [\apj] {10.3847/1538-4357/ab846a}, \href
  {https://ui.adsabs.harvard.edu/abs/2020ApJ...893...48N} {893, 48}

\bibitem[\protect\citeauthoryear{{Naidu} et~al.,}{{Naidu}
  et~al.}{2022}]{Naidu2022}
{Naidu} R.~P.,  et~al., 2022, \mn@doi [\apjl] {10.3847/2041-8213/ac5589}, \href
  {https://ui.adsabs.harvard.edu/abs/2022ApJ...926L..36N} {926, L36}

\bibitem[\protect\citeauthoryear{{Norris}, {van de Ven}, {Kannappan},
  {Schinnerer}  \& {Leaman}}{{Norris} et~al.}{2019}]{Norris.etal.2019}
{Norris} M.~A.,  {van de Ven} G.,  {Kannappan} S.~J.,  {Schinnerer} E.,
  {Leaman} R.,  2019, \mn@doi [\mnras] {10.1093/mnras/stz2096}, \href
  {https://ui.adsabs.harvard.edu/abs/2019MNRAS.488.5400N} {488, 5400}

\bibitem[\protect\citeauthoryear{{Oesch} et~al.,}{{Oesch}
  et~al.}{2016}]{Oesch2016}
{Oesch} P.~A.,  et~al., 2016, \mn@doi [\apj] {10.3847/0004-637X/819/2/129},
  \href {https://ui.adsabs.harvard.edu/abs/2016ApJ...819..129O} {819, 129}

\bibitem[\protect\citeauthoryear{Oliphant}{Oliphant}{2015}]{NumPy}
Oliphant T.~E.,  2015, Guide to NumPy, 2nd edn.
CreateSpace Independent Publishing Platform, USA

\bibitem[\protect\citeauthoryear{{Ono} et~al.,}{{Ono} et~al.}{2022}]{Ono2022}
{Ono} Y.,  et~al., 2022, \mn@doi [arXiv e-prints] {10.48550/arXiv.2208.13582},
  \href {https://ui.adsabs.harvard.edu/abs/2022arXiv220813582O} {p.
  arXiv:2208.13582}

\bibitem[\protect\citeauthoryear{Pedregosa et~al.,}{Pedregosa
  et~al.}{2011}]{sklearn}
Pedregosa F.,  et~al., 2011, J. Mach. Learn. Res., 12, 2825–2830

\bibitem[\protect\citeauthoryear{{P{\'e}rez-Villegas}, {Barbuy}, {Kerber},
  {Ortolani}, {Souza}  \& {Bica}}{{P{\'e}rez-Villegas}
  et~al.}{2020}]{Perez2020}
{P{\'e}rez-Villegas} A.,  {Barbuy} B.,  {Kerber} L.~O.,  {Ortolani} S.,
  {Souza} S.~O.,   {Bica} E.,  2020, \mn@doi [\mnras] {10.1093/mnras/stz3162},
  \href {https://ui.adsabs.harvard.edu/abs/2020MNRAS.491.3251P} {491, 3251}

\bibitem[\protect\citeauthoryear{{Phillips} et~al.,}{{Phillips}
  et~al.}{2022}]{Phillips2022}
{Phillips} S.~G.,  et~al., 2022, \mn@doi [\mnras] {10.1093/mnras/stab3532},
  \href {https://ui.adsabs.harvard.edu/abs/2022MNRAS.510.3727P} {510, 3727}

\bibitem[\protect\citeauthoryear{{Read} \& {Erkal}}{{Read} \&
  {Erkal}}{2019}]{Read.Erkal.2019}
{Read} J.~I.,  {Erkal} D.,  2019, \mn@doi [\mnras] {10.1093/mnras/stz1320},
  \href {https://ui.adsabs.harvard.edu/abs/2019MNRAS.487.5799R} {487, 5799}

\bibitem[\protect\citeauthoryear{{Ricotti}}{{Ricotti}}{2002}]{Ricotti.2002}
{Ricotti} M.,  2002, \mn@doi [\mnras] {10.1046/j.1365-8711.2002.05990.x}, \href
  {https://ui.adsabs.harvard.edu/abs/2002MNRAS.336L..33R} {336, L33}

\bibitem[\protect\citeauthoryear{{Rix} et~al.,}{{Rix} et~al.}{2022}]{Rix2022}
{Rix} H.-W.,  et~al., 2022, \mn@doi [\apj] {10.3847/1538-4357/ac9e01}, \href
  {https://ui.adsabs.harvard.edu/abs/2022ApJ...941...45R} {941, 45}

\bibitem[\protect\citeauthoryear{{Robertson} et~al.,}{{Robertson}
  et~al.}{2023}]{Robertson2023}
{Robertson} B.~E.,  et~al., 2023, \mn@doi [Nature Astronomy]
  {10.1038/s41550-023-01921-1}, \href
  {https://ui.adsabs.harvard.edu/abs/2023NatAs.tmp...67R} {}

\bibitem[\protect\citeauthoryear{{Rodriguez}, {Hafen}, {Grudi{\'c}},
  {Lamberts}, {Sharma}, {Faucher-Gigu{\`e}re}  \& {Wetzel}}{{Rodriguez}
  et~al.}{2023}]{Rodriguez2023}
{Rodriguez} C.~L.,  {Hafen} Z.,  {Grudi{\'c}} M.~Y.,  {Lamberts} A.,  {Sharma}
  K.,  {Faucher-Gigu{\`e}re} C.-A.,   {Wetzel} A.,  2023, \mn@doi [\mnras]
  {10.1093/mnras/stad578}, \href
  {https://ui.adsabs.harvard.edu/abs/2023MNRAS.521..124R} {521, 124}

\bibitem[\protect\citeauthoryear{{Sanders}, {Belokurov}  \& {Man}}{{Sanders}
  et~al.}{2021}]{Sanders2021}
{Sanders} J.~L.,  {Belokurov} V.,   {Man} K. T.~F.,  2021, \mn@doi [\mnras]
  {10.1093/mnras/stab1951}, \href
  {https://ui.adsabs.harvard.edu/abs/2021MNRAS.506.4321S} {506, 4321}

\bibitem[\protect\citeauthoryear{{Schechter}}{{Schechter}}{1976}]{Schechter1976}
{Schechter} P.,  1976, \mn@doi [\apj] {10.1086/154079}, \href
  {https://ui.adsabs.harvard.edu/abs/1976ApJ...203..297S} {203, 297}

\bibitem[\protect\citeauthoryear{{Schiavon} et~al.,}{{Schiavon}
  et~al.}{2017}]{Schiavon2017}
{Schiavon} R.~P.,  et~al., 2017, \mn@doi [\mnras] {10.1093/mnras/stw2162},
  \href {https://ui.adsabs.harvard.edu/abs/2017MNRAS.465..501S} {465, 501}

\bibitem[\protect\citeauthoryear{{Senchyna}, {Plat}, {Stark}  \&
  {Rudie}}{{Senchyna} et~al.}{2023}]{Senchyna2023}
{Senchyna} P.,  {Plat} A.,  {Stark} D.~P.,   {Rudie} G.~C.,  2023, \mn@doi
  [arXiv e-prints] {10.48550/arXiv.2303.04179}, \href
  {https://ui.adsabs.harvard.edu/abs/2023arXiv230304179S} {p. arXiv:2303.04179}

\bibitem[\protect\citeauthoryear{{Spitler} \& {Forbes}}{{Spitler} \&
  {Forbes}}{2009}]{Spitler.Forbes.2009}
{Spitler} L.~R.,  {Forbes} D.~A.,  2009, \mn@doi [\mnras]
  {10.1111/j.1745-3933.2008.00567.x}, \href
  {https://ui.adsabs.harvard.edu/abs/2009MNRAS.392L...1S} {392, L1}

\bibitem[\protect\citeauthoryear{{Tacchella} et~al.,}{{Tacchella}
  et~al.}{2023}]{Tacchella2023}
{Tacchella} S.,  et~al., 2023, \mn@doi [arXiv e-prints]
  {10.48550/arXiv.2302.07234}, \href
  {https://ui.adsabs.harvard.edu/abs/2023arXiv230207234T} {p. arXiv:2302.07234}

\bibitem[\protect\citeauthoryear{{Tang}, {Liu}, {Fern{\'a}ndez-Trincado},
  {Geisler}, {Shi}, {Zamora}, {Worthey}  \& {Moreno}}{{Tang}
  et~al.}{2019}]{Tang2019}
{Tang} B.,  {Liu} C.,  {Fern{\'a}ndez-Trincado} J.~G.,  {Geisler} D.,  {Shi}
  J.,  {Zamora} O.,  {Worthey} G.,   {Moreno} E.,  2019, \mn@doi [\apj]
  {10.3847/1538-4357/aaf6b1}, \href
  {https://ui.adsabs.harvard.edu/abs/2019ApJ...871...58T} {871, 58}

\bibitem[\protect\citeauthoryear{{Tang}, {Fern{\'a}ndez-Trincado}, {Liu}, {Yu},
  {Yan}, {Gao}, {Shi}  \& {Geisler}}{{Tang} et~al.}{2020}]{Tang2020}
{Tang} B.,  {Fern{\'a}ndez-Trincado} J.~G.,  {Liu} C.,  {Yu} J.,  {Yan} H.,
  {Gao} Q.,  {Shi} J.,   {Geisler} D.,  2020, \mn@doi [\apj]
  {10.3847/1538-4357/ab7233}, \href
  {https://ui.adsabs.harvard.edu/abs/2020ApJ...891...28T} {891, 28}

\bibitem[\protect\citeauthoryear{{VandenBerg}, {Brogaard}, {Leaman}  \&
  {Casagrande}}{{VandenBerg} et~al.}{2013}]{VdB2013}
{VandenBerg} D.~A.,  {Brogaard} K.,  {Leaman} R.,   {Casagrande} L.,  2013,
  \mn@doi [\apj] {10.1088/0004-637X/775/2/134}, \href
  {https://ui.adsabs.harvard.edu/abs/2013ApJ...775..134V} {775, 134}

\bibitem[\protect\citeauthoryear{{Vanzella} et~al.,}{{Vanzella}
  et~al.}{2023}]{Vanzella.etal.2023}
{Vanzella} E.,  et~al., 2023, \mn@doi [\apj] {10.3847/1538-4357/acb59a}, \href
  {https://ui.adsabs.harvard.edu/abs/2023ApJ...945...53V} {945, 53}

\bibitem[\protect\citeauthoryear{{Vasiliev} \& {Baumgardt}}{{Vasiliev} \&
  {Baumgardt}}{2021}]{Vasiliev2021}
{Vasiliev} E.,  {Baumgardt} H.,  2021, \mn@doi [\mnras]
  {10.1093/mnras/stab1475}, \href
  {https://ui.adsabs.harvard.edu/abs/2021MNRAS.505.5978V} {505, 5978}

\bibitem[\protect\citeauthoryear{{Vasiliev}, {Belokurov}  \&
  {Evans}}{{Vasiliev} et~al.}{2022}]{radialize}
{Vasiliev} E.,  {Belokurov} V.,   {Evans} N.~W.,  2022, \mn@doi [\apj]
  {10.3847/1538-4357/ac4fbc}, \href
  {https://ui.adsabs.harvard.edu/abs/2022ApJ...926..203V} {926, 203}

\bibitem[\protect\citeauthoryear{{Vogelsberger}, {Marinacci}, {Torrey}  \&
  {Puchwein}}{{Vogelsberger} et~al.}{2020}]{Vogelsberger.etal.2020}
{Vogelsberger} M.,  {Marinacci} F.,  {Torrey} P.,   {Puchwein} E.,  2020,
  \mn@doi [Nature Reviews Physics] {10.1038/s42254-019-0127-2}, \href
  {https://ui.adsabs.harvard.edu/abs/2020NatRP...2...42V} {2, 42}

\bibitem[\protect\citeauthoryear{{Weinberg}, {Andrews}  \&
  {Freudenburg}}{{Weinberg} et~al.}{2017}]{Weinberg2017}
{Weinberg} D.~H.,  {Andrews} B.~H.,   {Freudenburg} J.,  2017, \mn@doi [\apj]
  {10.3847/1538-4357/837/2/183}, \href
  {https://ui.adsabs.harvard.edu/abs/2017ApJ...837..183W} {837, 183}

\bibitem[\protect\citeauthoryear{{Weisz}, {Savino}  \& {Dolphin}}{{Weisz}
  et~al.}{2023}]{Weisz.etal.2023}
{Weisz} D.~R.,  {Savino} A.,   {Dolphin} A.~E.,  2023, \mn@doi [\apj]
  {10.3847/1538-4357/acc328}, \href
  {https://ui.adsabs.harvard.edu/abs/2023ApJ...948...50W} {948, 50}

\bibitem[\protect\citeauthoryear{{Wetzel} et~al.,}{{Wetzel}
  et~al.}{2023}]{Wetzel.etal.2023}
{Wetzel} A.,  et~al., 2023, \mn@doi [\apjs] {10.3847/1538-4365/acb99a}, \href
  {https://ui.adsabs.harvard.edu/abs/2023ApJS..265...44W} {265, 44}

\bibitem[\protect\citeauthoryear{{Wilkins} et~al.,}{{Wilkins}
  et~al.}{2023}]{Wilkins.etal.2023}
{Wilkins} S.~M.,  et~al., 2023, \mn@doi [\mnras] {10.1093/mnras/stac3280},
  \href {https://ui.adsabs.harvard.edu/abs/2023MNRAS.519.3118W} {519, 3118}

\bibitem[\protect\citeauthoryear{{Yung}, {Somerville}, {Finkelstein}, {Wilkins}
   \& {Gardner}}{{Yung} et~al.}{2023}]{Yung.etal.2023}
{Yung} L.~Y.~A.,  {Somerville} R.~S.,  {Finkelstein} S.~L.,  {Wilkins} S.~M.,
  {Gardner} J.~P.,  2023, \mn@doi [arXiv e-prints] {10.48550/arXiv.2304.04348},
  \href {https://ui.adsabs.harvard.edu/abs/2023arXiv230404348Y} {p.
  arXiv:2304.04348}

\bibitem[\protect\citeauthoryear{{Zick}, {Weisz}  \& {Boylan-Kolchin}}{{Zick}
  et~al.}{2018}]{Zick.etal.2018}
{Zick} T.~O.,  {Weisz} D.~R.,   {Boylan-Kolchin} M.,  2018, \mn@doi [\mnras]
  {10.1093/mnras/sty662}, \href
  {https://ui.adsabs.harvard.edu/abs/2018MNRAS.477..480Z} {477, 480}

\makeatother
\end{thebibliography}

\appendix

\label{lastpage}

\end{document}